\documentclass[aps,twocolumn,amsmath,amssymb]{revtex4}

\usepackage[latin1]{inputenc}
\usepackage{bm}
\usepackage{xcolor}
\usepackage{tikz}
\usepackage{graphicx}
\usepackage{subfig}
\usepackage{multirow}
\usepackage{rotating}
\usetikzlibrary{trees,arrows,decorations.pathmorphing}
\usepackage{array}
\newcolumntype{H}{>{\setbox0=\hbox\bgroup}c<{\egroup}@{}}
\newcolumntype{G}{@{}>{\lrbox0}l<{\endlrbox}}

\renewcommand{\b}[1]{\ensuremath{\mathbf{#1}}}
\newcommand{\la}{\langle}
\newcommand{\ra}{\rangle}

\newcommand{\mx}{\mathbf{x}}
\newcommand{\mr}{\mathbf{r}}
\newcommand{\mA}{\mathbf{A}}
\newcommand{\mB}{\mathbf{B}}
\newcommand{\mK}{\mathbf{K}}
\newcommand{\mX}{\mathbf{X}}
\newcommand{\mY}{\mathbf{Y}}

\newcommand{\fl}{\rightarrow}
\newcommand{\erf}{\mathrm{erf}}
\newcommand{\RSH}{\text{RSH} }
\newcommand{\fHx}{\text{fHx}}
\newcommand{\HF}{\text{HF}}
\newcommand{\LDA}{\text{LDA}}

\newcommand{\up}{\uparrow}
\newcommand{\down}{\downarrow}
\renewcommand{\H}{\text{H}}
\newcommand{\LC}{\text{LC} }
\newcommand{\beq}{\begin{equation}}
\newcommand{\eeq}{\end{equation}}

\begin{document}
\title{Electronic excitations from a linear-response range-separated hybrid scheme}
\author{Elisa Rebolini}\email{rebolini@lct.jussieu.fr}
\author{Andreas Savin}\email{savin@lct.jussieu.fr}
\author{Julien Toulouse}\email{julien.toulouse@upmc.fr}
\affiliation{Laboratoire de Chimie Théorique, Université Pierre et Marie Curie and CNRS, 75005 Paris, France}

\date{\today}

\begin{abstract}
We study linear-response time-dependent density-functional theory (DFT) based on the single-determinant range-separated hybrid (RSH) scheme, i.e. combining a long-range Hartree-Fock exchange kernel with a short-range DFT exchange-correlation kernel, for calculating electronic excitation energies of molecular systems. It is an alternative to the more common long-range correction (LC) scheme which combines a long-range Hartree-Fock exchange kernel with a short-range DFT exchange kernel and a standard full-range DFT correlation kernel. We discuss the local-density approximation (LDA) to the short-range exchange and correlation kernels, and assess the performance of the linear-response RSH scheme for singlet $\to$ singlet and singlet $\to$ triplet valence and Rydberg excitations in the N$_2$, CO, H$_2$CO, C$_2$H$_4$, and C$_6$H$_6$ molecules, and for the first charge-transfer excitation in the C$_2$H$_4$-C$_2$F$_4$ dimer. For these systems, the presence of long-range LDA correlation in the ground-state calculation and in the linear-response kernel has only a small impact on the excitation energies and oscillator strengths, so that the RSH method gives results very similar to the ones given by the LC scheme. Like in the LC scheme, the introduction of long-range HF exchange in the present method corrects the underestimation of charge-transfer and high-lying Rydberg excitation energies obtained with standard (semi)local density-functional approximations, but also leads to underestimated excitation energies to low-lying spin-triplet valence states. This latter problem is largely cured by the Tamm-Dancoff approximation which leads to a relatively uniform accuracy for all excitation energies. This work thus suggests that the present linear-response RSH scheme is a reasonable starting approximation for describing electronic excitation energies, even before adding an explicit treatment of long-range correlation.
\end{abstract}

\maketitle

\section{Introduction}

Range-separated density-functional theory (see, e.g., Ref.~\onlinecite{TouColSav-PRA-04} and references therein) constitutes an alternative to standard Kohn-Sham (KS) density-functional theory (DFT)~\cite{KohSha-PR-65} for ground-state electronic-structure calculations. It consists in combining wave-function-type approximations for long-range electron-electron interactions with density-functional approximations for short-range electron-electron interactions, using a controllable range-separation parameter. For example, in the single-determinant range-separated hybrid (RSH) scheme~\cite{AngGerSavTou-PRA-05}, the long-range Hartree-Fock (HF) exchange energy is combined with a short-range exchange-correlation density-functional approximation. The long-range correlation energy is missing in this scheme, but it can be added in a second step by many-body perturbation theory for describing van der Waals dispersion interactions for instance~\cite{AngGerSavTou-PRA-05,GolWerSto-PCCP-05,TouGerJanSavAng-PRL-09,JanHenScu-JCP-09,TouZhuAngSav-PRA-10,TouZhuSavJanAng-JCP-11}. A simpler approach is the long-range correction (LC) scheme~\cite{IikTsuYanHir-JCP-01}, also called RSHX~\cite{GerAng-CPL-05a}, which consists in applying range separation on exchange only, i.e. combining the long-range HF exchange energy with a short-range exchange density-functional approximation and using a standard full-range correlation density functional. More complicated decompositions of the exchange energy have also been proposed, such as in the CAM-B3LYP approximation~\cite{YanTewHan-CPL-04}. 

Range separation is also applied in linear-response time-dependent density-functional theory (TDDFT)~\cite{GroKoh-PRL-85} for calculating excitation energies and other response properties. The first and probably most widely used range-separated TDDFT approach is based on the LC scheme~\cite{TawTsuYanYanHir-JCP-04}, and involves a long-range HF exchange kernel combined with a short-range DFT exchange kernel and a standard full-range DFT correlation kernel. It has also been proposed to use in this scheme an empirically modified correlation density functional depending on the range-separation parameter~\cite{LivBae-PCCP-07}. The CAM-B3LYP scheme and other similar schemes have also been applied in linear-response theory for calculating excitation energies~\cite{YanTewHan-CPL-04,YanHarHan-MP-05,PeaHelSalKeaLutTozHan-PCCP-06,PeaCohToz-PCCP-06,ChaHea-JCP-08,LanRohHer-JPCB-08,RohHer-JCP-08,AkiTen-CPL-08,RohMarHer-JCP-09,AkiTen-IJQC-09,PevTru-JPCL-11,NguDayPac-JCP-11,LinTsaLiCha-JCP-12}. In all these schemes, the presence of long-range HF exchange greatly improves Rydberg and charge-transfer excitation energies, in comparison to time-dependent Kohn-Sham (TDKS) calculations using standard local or semilocal density-functional approximations in which they are strongly underestimated (see, e.g., Ref.~\onlinecite{PeaBenHelToz-JCP-08}).

In this paper, we study a range-separated linear-response TDDFT method based on the RSH scheme, i.e. combining a long-range HF exchange kernel with a short-range DFT exchange-correlation kernel with no long-range correlation kernel. The motivation for this range-separated TDDFT approach is that, as for exchange, the long-range part of standard correlation density-functional approximations such as the local-density approximation (LDA) is usually inaccurate~\cite{TouColSav-PRA-04,TouColSav-JCP-05,TouColSav-MP-05}, so one may as well remove it. This can be viewed as a first-level approximation before adding a more accurate treatment of long-range correlation, e.g., by linear-response density-matrix functional theory (DMFT)~\cite{Per-JCP-12} or linear-response multiconfiguration self-consistent field (MCSCF) theory~\cite{FroKneJen-JCP-13}. These last approaches are capable of describing excited states of double excitation character, which are out of reach within a single-determinant linear-response scheme using adiabatic exchange-correlation kernels (except in a spin-flip formulation~\cite{WanZie-JCP-04,CasHui-ARPC-12}). 

The main goal of this paper is to test whether the range-separated TDDFT method based on the RSH scheme is a reasonable starting approximation for calculating excitation energies of molecular systems, even before adding explicit long-range correlations. For this purpose, we apply the method to singlet $\to$ singlet and singlet $\to$ triplet valence and Rydberg excitations in the N$_2$, CO, H$_2$CO, C$_2$H$_4$, and C$_6$H$_6$ molecules, and to the first charge-transfer (CT) excitation in the C$_2$H$_4$-C$_2$F$_4$ dimer, and compare with the LC scheme, as well as non-range-separated methods. In particular, we study the effect of dropping long-range LDA correlation in comparison to the LC scheme.

The paper is organized as follows. The working equations of the linear-response RSH scheme are laid down in Section~\ref{sec:rsTDDFT}, and the short-range DFT exchange and correlation kernels are discussed in Section~\ref{sec:srkernels}. After giving computational details in Section~\ref{sec:details}, we report and discuss our results in Section~\ref{sec:results}. Section~\ref{sec:conclusion} summarizes our conclusions. Technical details are given in Appendices. Hartree atomic units are assumed throughout unless otherwise indicated.

\section{Linear-response range-separated hybrid scheme}
\label{sec:rsTDDFT}

\subsection{Ground-state range-separated scheme}

In the RSH scheme~\cite{AngGerSavTou-PRA-05}, the ground-state energy is approximated as the following minimum over single-determinant wave functions $\Phi$,
\beq
\begin{split}
E_{\RSH}  = &  \min_{\Phi} \lbrace \la \Phi | \hat{T}  + \hat{V}_{ext} | \Phi \ra 
			 + E_\H[n_\Phi] \\ &+ E_{x,\HF}^{lr}[\Phi] + E_{xc}^{sr}[n_\Phi,m_\Phi] \rbrace,
\label{RSH}
\end{split}
\eeq
where $\hat{T}$ is the kinetic energy operator, $\hat{V}_{ext}$ is the external potential operator, $E_\H[n]$ is the Hartree energy density functional,
\beq E_\H[n] = \dfrac{1}{2} \int n(\mr_1)n(\mr_2)w_{ee}(|\mr_1 - \mr_2|)d\mr_1 d\mr_2, \eeq
with the Coulombic electron-electron interaction $w_{ee}(|\mr_1 - \mr_2|) =1/|\mr_1 - \mr_2|$, $E_{x,\HF}^{lr}[\Phi]$ is the long-range HF exchange energy
\beq E_{x,\HF}^{lr}[\Phi] =-\dfrac{1}{2} \int | \la \Phi |\hat{n}_1(\mx_1,\mx_2)|\Phi \ra |^2 w_{ee}^{lr}(|\mr_1 - \mr_2|) d\mx_1 d\mx_2, \eeq 
with the one-particle density-matrix operator $\hat{n}_1(\mx_1,\mx_2)$ and a long-range electron-electron interaction $w_{ee}^{lr}(|\mr_1 - \mr_2|)$, and $E_{xc}^{sr}[n,m]$ is the short-range exchange-correlation energy functional depending on the total density $n(\mr) = n_\uparrow(\mr) + n_\downarrow(\mr)$ and the (collinear) spin magnetization density $m(\mr) = n_\uparrow(\mr) - n_\downarrow(\mr)$, written with the spin densities $n_\sigma(\mr) = n(\mx)$ for the space-spin coordinate $\mx=(\mr,\sigma)$.
In this work, the long-range interaction will be taken as $w_{ee}^{lr}(r) = \erf(\mu r)/r$, where the parameter $\mu$ can be interpreted as the inverse of a smooth ``cut-off'' radius, but other interactions have also been considered~\cite{SavFla-IJQC-95,Sav-INC-96a,BaeNeu-PRL-05}. What is neglected in Eq.~(\ref{RSH}) is the long-range correlation energy $E_c^{lr}$, but it can be added \textit{a posteriori} by perturbative methods \cite{AngGerSavTou-PRA-05,GolWerSto-PCCP-05,FroJen-PRA-08,TouGerJanSavAng-PRL-09,JanHenScu-JCP-09,PaiJanHenScuGruKre-JCP-10,TouZhuAngSav-PRA-10,TouZhuSavJanAng-JCP-11}.

In the LC scheme~\cite{IikTsuYanHir-JCP-01}, range separation is applied to the exchange energy only and the ground-state energy is expressed as
\beq
\begin{split}
E_{\LC}  = &  \min_{\Phi} \lbrace \la \Phi | \hat{T}  + \hat{V}_{ext} | \Phi \ra 
			 + E_\H[n_\Phi] \\ &+ E_{x,\HF}^{lr}[\Phi] + E_{x}^{sr}[n_\Phi,m_\Phi] + E_{c}[n_\Phi,m_\Phi] \rbrace,
\end{split}
\eeq
where $E_{x}^{sr}[n,m]$ is the short-range exchange energy functional, and $E_{c}[n,m]$ is the full-range correlation energy functional.

\subsection{Linear-response theory}

Just like in standard TDDFT~\cite{GroKoh-PRL-85}, time-dependent linear-response theory applied to the RSH scheme leads to a familiar Dyson-like equation for the frequency-dependent 4-point linear response function $\chi(\mx_1,\mx_2;\mx_1',\mx_2';\omega)$ to a time-dependent perturbation (dropping the space-spin coordinates for simplicity)
\beq \label{eq:dyson_chi}
\chi^{-1}(\omega) = \chi^{-1}_0(\omega) - f_H - f_{x,\HF}^{lr} - f_{xc}^{sr},
\eeq
where $\chi_0(\omega)$ is the non-interacting \RSH response function, $f_H$ is the Hartree kernel,
\beq
f_H(\mx_1,\mx_2;\mx_1',\mx_2') = w_{ee}(|\mr_1-\mr_2|) \delta(\mx_1-\mx_1') \delta(\mx_2-\mx_2'),
\eeq
$f_{x,\HF}^{lr}$ is the long-range HF exchange kernel,
\beq
f_{x,\HF}^{lr}(\mx_1,\mx_2;\mx_1',\mx_2') = -w_{ee}^{lr}(|\mr_1-\mr_2|)\delta(\mx_1-\mx_2') \delta(\mx_1'-\mx_2), 
\eeq
and $f_{xc}^{sr}$ is the short-range exchange-correlation kernel which is frequency independent in the adiabatic approximation,
\beq
f_{xc}^{sr}(\mx_1,\mx_2;\mx_1',\mx_2') = f_{xc}^{sr}(\mx_1,\mx_2) \delta(\mx_1-\mx_1') \delta(\mx_2-\mx_2'),
\eeq
with the 2-point kernel
\beq
f_{xc}^{sr}(\mx_1,\mx_2) = \dfrac{\delta^2 E_{xc}^{sr}[n,m]}{\delta n(\mx_1) \delta n(\mx_2)}.
\eeq
Note that a 4-point formalism is required here because of the HF exchange kernel.
The excitation energies are given by the poles of $\chi(\omega)$ in $\omega$. Working in the basis of the RSH spin orbitals $\{\phi_k(\mx)\}$, the poles can be found by the pseudo-Hermitian eigenvalue problem~\cite{Cas-INC-95}
\beq\label{eq:spin-orbital casida}
\begin{pmatrix}
\mA & \mB \\ 
\mB^* & \mA^*
\end{pmatrix} 
\begin{pmatrix}
\mX_n \\ \mY_n
\end{pmatrix} = \omega_n
\begin{pmatrix}
\bm{1} & \bm{0} \\
\bm{0} & \bm{-1}
\end{pmatrix}
\begin{pmatrix}
\mX_n \\ \mY_n
\end{pmatrix},
\eeq
whose solutions come in pairs: the excitation energy $\omega_n$ associated with the eigenvector $(\mX_n,\mY_n)$, and the deexcitation energy $-\omega_n$ associated with $(\mY_n^*,\mX_n^*)$. The matrix elements of $\mA$ and $\mB$ are
\beq
\begin{split}
\mA_{ia,jb} &= (\varepsilon_a - \varepsilon_i) \delta_{ij} \delta_{ab} + \mathbf{K}_{ia, jb}, \\
\mB_{ia, jb} & = \mathbf{K}_{ia, bj},
\end{split}
\eeq
where $i,j$ and $a,b$ refer to occupied and virtual spin orbitals, respectively, $\varepsilon_k$ is the energy of the spin orbital $k$, and $\mathbf{K}$ is the coupling matrix accounting for the contributions of the different kernels,
\beq 
\begin{split}
\mathbf{K}_{ia,jb}  
& = \la aj| \hat{f}_H |ib \ra + \la aj|\hat{f}_{x,\HF}^{lr}|ib \ra + \la aj |\hat{f}_{xc}^{sr} |ib \ra \\
& = \la aj| \hat{w}_{ee} |ib \ra  - \la aj|\hat{w}_{ee}^{lr}|bi \ra  + \la aj |\hat{f}_{xc}^{sr} |ib \ra,
\end{split}
\label{eq:Kiajb}
\eeq
where $\la aj| \hat{w}_{ee} |ib \ra$ and $\la aj|\hat{w}_{ee}^{lr}|bi \ra$ are the two-electron integrals for the Coulombic and long-range interactions, respectively, and $\la aj |\hat{f}_{xc}^{sr} |ib \ra$ are the matrix elements of the short-range exchange-correlation kernel,
\beq 
\begin{split}
\la aj |\hat{f}_{xc}^{sr} |ib \ra &= \int \phi_a^*(\mx_1) \phi_j^*(\mx_2) f_{xc}^{sr}(\mx_1,\mx_2) \\
&\times\phi_i(\mx_1) \phi_b(\mx_2) d\mx_1 d \mx_2.
\end{split}
\eeq
For real-valued orbitals, and if $\b{A}+\b{B}$ and $\b{A}-\b{B}$ are positive definite, Eq.~(\ref{eq:spin-orbital casida}) is conveniently transformed into a half-size symmetric eigenvalue equation~\cite{Cas-INC-95}
\begin{eqnarray} 
\b{M} \, \b{Z}_{n} = \omega_{n}^2 \, \b{Z}_{n},
\label{MZ}
\end{eqnarray}
with $\b{M}=\left( \b{A}-\b{B} \right)^{1/2} \left( \b{A}+\b{B} \right) \left( \b{A}-\b{B} \right)^{1/2}$ and the normalized eigenvectors $\b{Z}_{n} = \sqrt{\omega_{n}} \left( \b{A}-\b{B} \right)^{-1/2} \left( \b{X}_{n}+\b{Y}_{n} \right)$. The Tamm-Dancoff approximation (TDA)~\cite{HirHea-CPL-99} consists in neglecting the coupling between the excitations and the de-excitations, i.e. setting $\mB=\bm{0}$. We note, in passing, that the TDA can also be viewed as a non-self-consistent approximation to the static (multiplet-sum) $\Delta$SCF method, which identifies the excited states with stationary points on the ground-state energy surface as a function of the orbital parameters~\cite{ZieSetKryAutWan-JCP-09,CasHui-ARPC-12}.

The same equations apply identically to the {\LC} scheme except that the short-range correlation kernel $f_c^{sr}$ has to be replaced by the full-range one $f_c$~\cite{TawTsuYanYanHir-JCP-04}.

\subsection{Spin adaptation for closed-shell systems}

For spin-restricted closed-shell calculations, Eq.~(\ref{MZ}) can be decoupled into two independent eigenvalue equations for singlet $\to$ singlet excitations and for singlet $\to$ triplet excitations, respectively~\cite{PetGro-IJQC-96,BauAhl-CPL-96,GisSniBae-CPC-99} (see Appendix~\ref{appendix:spin-kernel}). For simplicity, they will be referred to as ``singlet excitations'' and ``triplet excitations''. The modifications for spin adaptation are located in the expression of the coupling matrix $\mathbf{K}$, which becomes, for singlet excitations,
\beq
\begin{split}
^1 \mK_{ia,jb} 	& = 2 \la aj| \hat{w}_{ee} |ib \ra  - \la aj|\hat{w}_{ee}^{lr}|bi \ra + 2 \la aj |\,^1\hat{f}_{xc}^{sr} |ib \ra,
\end{split}
\label{eq:1Kiajb}
\eeq
and, for triplet excitations,
\beq
\begin{split}
^3 \mK_{ia,jb} 	& = - \la aj|\hat{w}_{ee}^{lr}|bi \ra + 2 \la aj |\,^3\hat{f}_{xc}^{sr} |ib \ra,
\end{split}
\label{eq:3Kiajb}
\eeq
where the indices $i,j,a,b$ refer now to spatial orbitals and the singlet and triplet short-range exchange-correlation kernels are 
\beq 
^1 f_{xc}^{sr}(\mr_1,\mr_2) = \dfrac{\delta ^2 E_{xc}^{sr}[n,m]}{\delta n(\mr_1) \delta n(\mr_2)},
\label{eq:1fxcsr}
\eeq 
and
\beq  
^3 f_{xc}^{sr}(\mr_1,\mr_2) = \dfrac{\delta ^2 E_{xc}^{sr}[n,m]}{\delta m(\mr_1) \delta m(\mr_2)},
\label{eq:3fxcsr}
\eeq
where the derivatives are taken at zero spin magnetization density, $m(\mr)=0$. Because the spin-dependent exchange functional $E_x^{sr}[n,m]$ is constructed from the spin-independent one $E_x^{sr}[n]=E_x^{sr}[n,m=0]$ via the spin-scaling relation~\cite{OliPer-PRA-79}, $E_x^{sr}[n,m] = \left(E_x^{sr}[2n_\uparrow] + E_x^{sr}[2n_\downarrow]\right)/2$, one can show that the singlet and triplet exchange kernels are identical, and, for closed-shell systems, can be written with the spin-independent functional,
\beq  
f_{x}^{sr}(\mr_1,\mr_2) = {^1}f_{x}^{sr}(\mr_1,\mr_2) = {^3}f_{x}^{sr}(\mr_1,\mr_2) = \dfrac{\delta ^2 E_{x}^{sr}[n]}{\delta n(\mr_1) \delta n(\mr_2)}.
\eeq
Therefore, contrary to the case of the correlation functional, the dependence on the spin magnetization density does not need to be considered in practice in the exchange functional for closed-shell systems.

The oscillator strength $f_n$ for state $n$ is zero for a triplet excitation, and it is calculated with the following formula for a singlet excitation (in the dipole length form)~\cite{Cas-INC-95}
\begin{equation}
f_n = \frac{4}{3} \sum_{\alpha=x,y,z} \left( \sum_{ia} d_{\alpha,ia} \left[ \left({^1}\b{A}-{^1}\b{B} \right)^{1/2} {^1}\b{Z}_n \right]_{ia} \right)^2,
\end{equation}
where $d_{\alpha,ia}=\int \phi_i(\b{r}) r_{\alpha} \phi_a(\b{r})d\b{r}$ is the $\alpha$-component of the transition dipole moment between the spatial occupied and virtual orbitals $\phi_i(\b{r})$ and $\phi_a(\b{r})$.

\section{Short-range adiabatic exchange-correlation kernels}
\label{sec:srkernels}

We will consider here the short-range adiabatic exchange and correlation kernels in the local-density approximation (LDA).

\subsection{Exchange kernel}
\begin{figure}[b]
\includegraphics[scale=0.85]{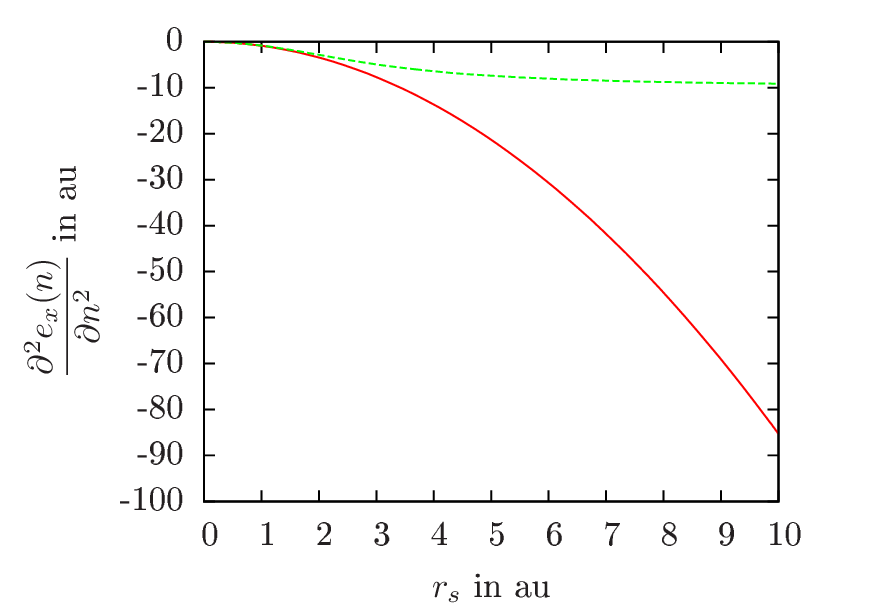}
\caption{Second-order derivatives of the full-range (full line) and short-range ($\mu=0.4$, dashed line) LDA exchange energy density with respect to the density $n$ as functions of the Wigner-Seitz radius $r_s$.}
\label{fig:fx}
\end{figure}
The short-range spin-independent LDA exchange energy functional is written as
\beq 
E_{x,\LDA}^{sr}[n] = \int e_{x}^{sr}(n(\mr)) d\mr,
\eeq
where $e_{x}^{sr}(n)=n \, \epsilon_{x}^{sr}(n)$ is the short-range energy density defined with the exchange energy per particle $\epsilon_{x}^{sr}(n)$ of the homogeneous electron gas (HEG) with the short-range electron-electron interaction $w_{ee}^{sr} = w_{ee} - w_{ee}^{lr}$. 
The analytic expression of $\epsilon_{x}^{sr}(n)$ is known~\cite{Sav-INC-96,TouSavFla-IJQC-04} and is recalled in Appendix~\ref{appendix:Exsr}. 
The short-range adiabatic LDA exchange kernel is given by the second-order derivative of the energy density with respect to the density,
\beq \begin{split}
f_{x,\LDA}^{sr}(\mr,\mr') & = \dfrac{\partial ^2 e_{x}^{sr}(n(\mr))}{\partial n^2}\,\delta(\mr-\mr').
\label{fxsrLDA}
\end{split} \eeq
Just like its full-range LDA counterpart, the short-range exchange LDA kernel is thus strictly local in space. However, this is here a less drastic approximation than for the full-range case. Indeed, by using the asymptotic expansion of the exact short-range spin-independent exchange density functional for $\mu\to\infty$~\cite{GilAdaPop-MP-96,TouColSav-PRA-04}, $E_x^{sr}[n] = -\pi/(4\mu^2)\int n(\mr)^2 d\mr + O\left(1/\mu^4\right)$, one can see that the {\it exact} adiabatic short-range exchange kernel has the following expansion in $1/\mu$,
\beq
f_{x}^{sr}(\mr,\mr') = -\frac{\pi}{2\mu^2} \, \delta(\mr-\mr') + O\left(\dfrac{1}{\mu^4}\right),
\label{fxsrmuinf}
\eeq
i.e., in the limit of a very short-range electron-electron interaction, it also becomes strictly local. More than that, the short-range LDA kernel of Eq.~(\ref{fxsrLDA}) is exact for the leading term of Eq.~(\ref{fxsrmuinf}), as shown in Appendix~\ref{appendix:Exsr}.

The short-range LDA exchange kernel for a fixed value of the range-separation parameter $\mu=0.4$ is shown in Fig.~\ref{fig:fx} as a function of the Wigner-Seitz radius $r_s= \left(3/(4\pi n)\right)^{1/3}$ and compared with the full-range LDA exchange kernel. 
The LDA exchange kernel is always negative, which is a consequence of the concavity of the LDA exchange energy density curve as a function of the density $n$. 
For high enough densities such that $r_s \ll 1/\mu$, the short-range LDA exchange kernel reduces to the full-range one (see Appendix~\ref{appendix:Exsr}). 
For larger values of $r_s$, the short-range LDA exchange kernel is reduced compared to the full-range one, and, in the low-density limit $r_s \to \infty$, it tends to the finite value of $-\pi/2\mu^2$ while the full-range LDA exchange kernel diverges to $-\infty$.

\subsection{Correlation kernel}
\begin{figure}[b]
\subfloat{\includegraphics[scale=0.85]{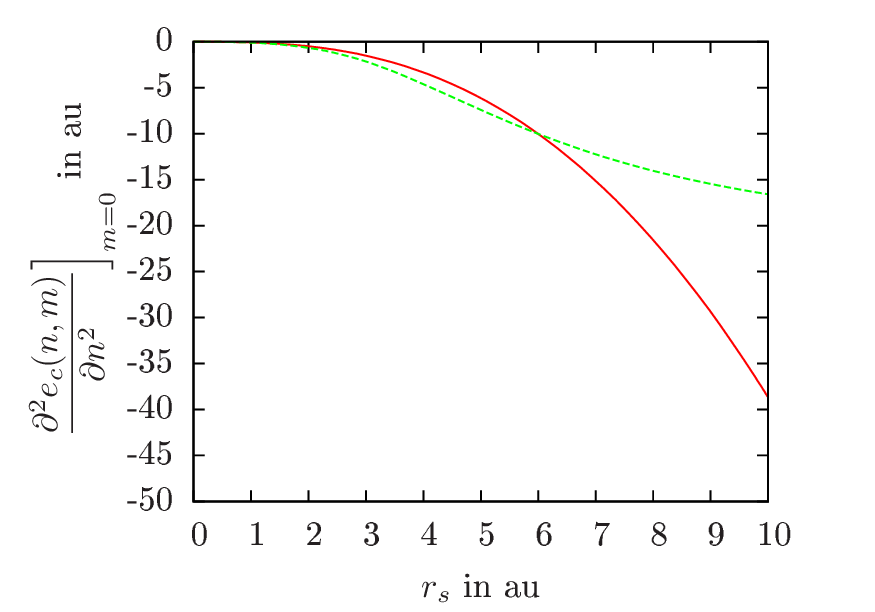}\label{fig:singfc}} \\
\subfloat{\includegraphics[scale=0.85]{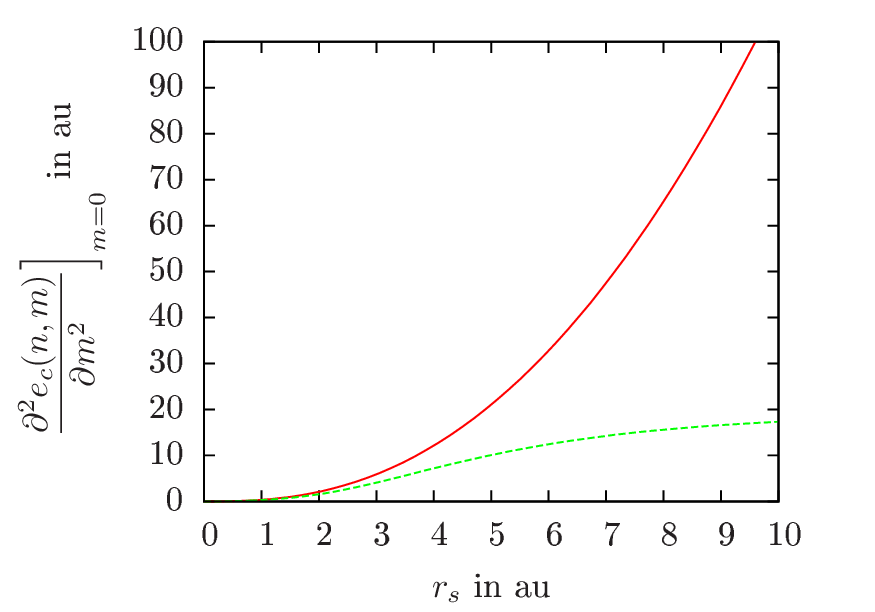}\label{fig:tripfc}}
\caption{Second-order derivatives of the full-range (full line) and short-range ($\mu=0.4$, dashed line) LDA correlation energy densities with respect to the density $n$ (top) and to the spin magnetization $m$ (bottom) evaluated at $m=0$ as functions of the Wigner-Seitz radius $r_s$. \label{fig:fc}}
\end{figure}

The short-range spin-dependent LDA correlation energy functional is written as
\beq 
E_{c,\LDA}^{sr}[n,m] = \int e_{c}^{sr}(n(\mr),m(\mr)) d\mr,
\eeq
where $e_{c}^{sr}(n,m)=n\,\epsilon_{c}(n,m)-n\,\epsilon_{c}^{lr}(n,m)$ is the complement short-range correlation energy density, obtained from the correlation energy per particle of the standard homogeneous electron gas (HEG), $\epsilon_{c}(n,m)$,~\cite{PerWan-PRB-92} and the correlation energy per particle of the HEG with the long-range electron-electron interaction, $\epsilon_{c}^{lr}(n,m)$, as parametrized from quantum Monte Carlo calculations by Paziani \textit{et al.}~\cite{PazMorGorBac-PRB-06}. Its expression is recalled in Appendix~\ref{appendix:Ecsr}. The singlet and triplet short-range adiabatic LDA correlation kernels are local functions given by the second-order derivatives of the energy density with respect to the density $n$ and the spin magnetization $m$, respectively,
\beq 
^1 f_{c,\LDA}^{sr}(\mr,\mr') = \dfrac{\partial ^2 e_{c}^{sr}(n(\mr),m(\mr))}{\partial n^2}\,\delta(\mr-\mr'),
\label{1fcLDAsr}
\eeq
\beq 
^3 f_{c,\LDA}^{sr}(\mr,\mr') = \dfrac{\partial ^2 e_{c}^{sr}(n(\mr),m(\mr))}{\partial m^2}\,\delta(\mr-\mr').
\label{3fcLDAsr}
\eeq
For closed-shell systems, these kernels need to be evaluated only at zero spin magnetization, $m=0$. Again, it can be argued that the strictly local form of the LDA correlation kernels of Eq.~(\ref{1fcLDAsr}) and~(\ref{3fcLDAsr}) is more appropriate for the short-range kernels than for the full-range ones. Using the asymptotic expansion of the exact short-range correlation functional for $\mu\to\infty$~\cite{TouColSav-PRA-04,GorSav-PRA-06}, $E_c^{sr}[n,m] = \pi/(2\mu^2)\int n_{2,c}(\mr,\mr) d\mr + 2\sqrt{2\pi}/(3\mu^2)\int n_{2}(\mr,\mr) d\mr + O\left(1/\mu^4\right)$, and the total and correlation on-top pair densities in the strong-interaction limit of the adiabatic connection $\lambda\to\infty$ (or for fully spin-polarized systems $n=|m|$)~\cite{BurPerErn-JCP-98,GorSeiSav-PCCP-08}, $n_{2}(\mr,\mr) \to 0$ and $n_{2,c}(\mr,\mr) \to -n(\mr)^2/2+m(\mr)^2/2$, it is easy to show that the leading terms in the expansions of the {\it exact} adiabatic short-range correlation kernels for $\mu\to\infty$, in the strong-interaction (or low-density) limit, are strictly local
\beq 
^1 f_{c}^{sr}(\mr,\mr') \xrightarrow[\lambda\to\infty]{} -\frac{\pi}{2\mu^2} \, \delta(\mr-\mr') + O\left(\dfrac{1}{\mu^4}\right),
\label{1fcr}
\eeq
\beq 
^3 f_{c}^{sr}(\mr,\mr') \xrightarrow[\lambda\to\infty]{} \frac{\pi}{2\mu^2} \, \delta(\mr-\mr') + O\left(\dfrac{1}{\mu^4}\right).
\label{3fcr}
\eeq
The short-range LDA correlation kernels of Eqs.~(\ref{1fcLDAsr}) and~(\ref{3fcLDAsr}), using the parametrization of Ref.~\onlinecite{PazMorGorBac-PRB-06}, are exact for these leading terms.

The singlet and triplet short-range LDA correlation kernels are plotted in Fig.~\ref{fig:fc}, and compared with the full-range LDA correlation kernels. 
The singlet LDA correlation kernel is always negative while the triplet LDA correlation kernel is always positive, 
reflecting the fact that the LDA correlation energy density is concave when plotted as a function of the density $n$ and convex when plotted as a function of the spin magnetization $m$. 
As for the exchange kernels, the singlet and triplet short-range LDA correlation kernels reduce to the full-range kernels in the high-density limit $r_s \to 0$ (see Appendix~\ref{appendix:Ecsr}).
In the low-density limit $r_s \to \infty$, they tend to the finite values of $\mp \pi/2\mu^2$, while the full-range kernels diverge to $\mp \infty$.

\section{Computational details}
\label{sec:details}

\begin{figure*}[floatfix]
\subfloat[Singlet]{\includegraphics[scale=0.85]{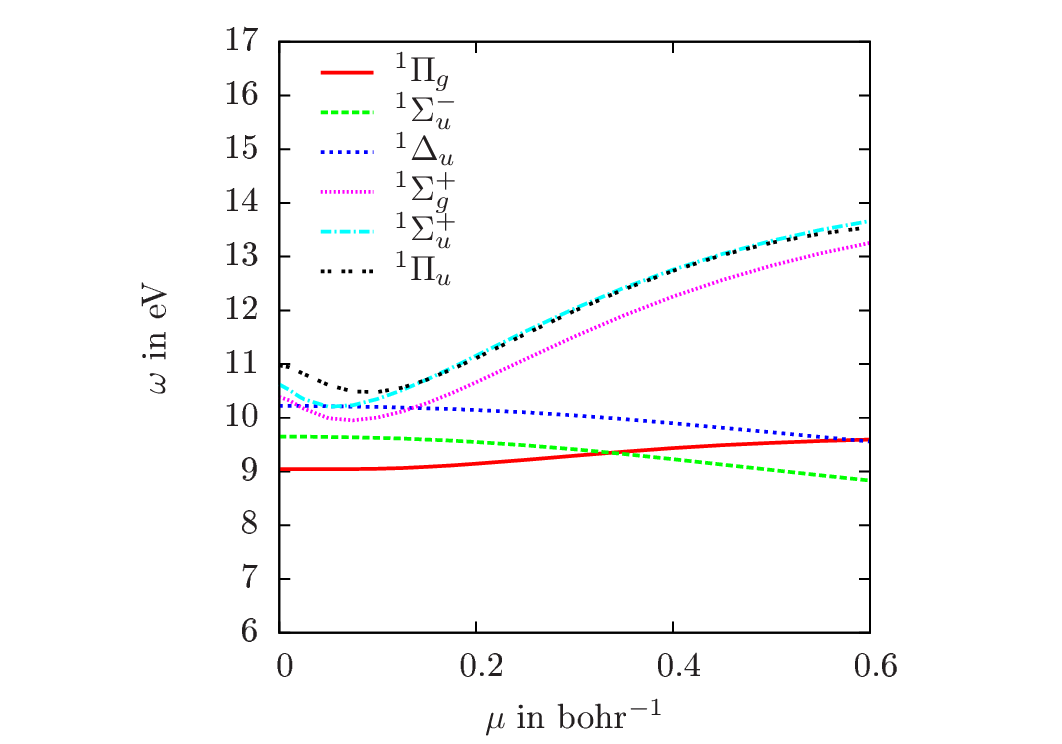}} 
\hspace*{-1.5cm}
\subfloat[Triplet]{\includegraphics[scale=0.85]{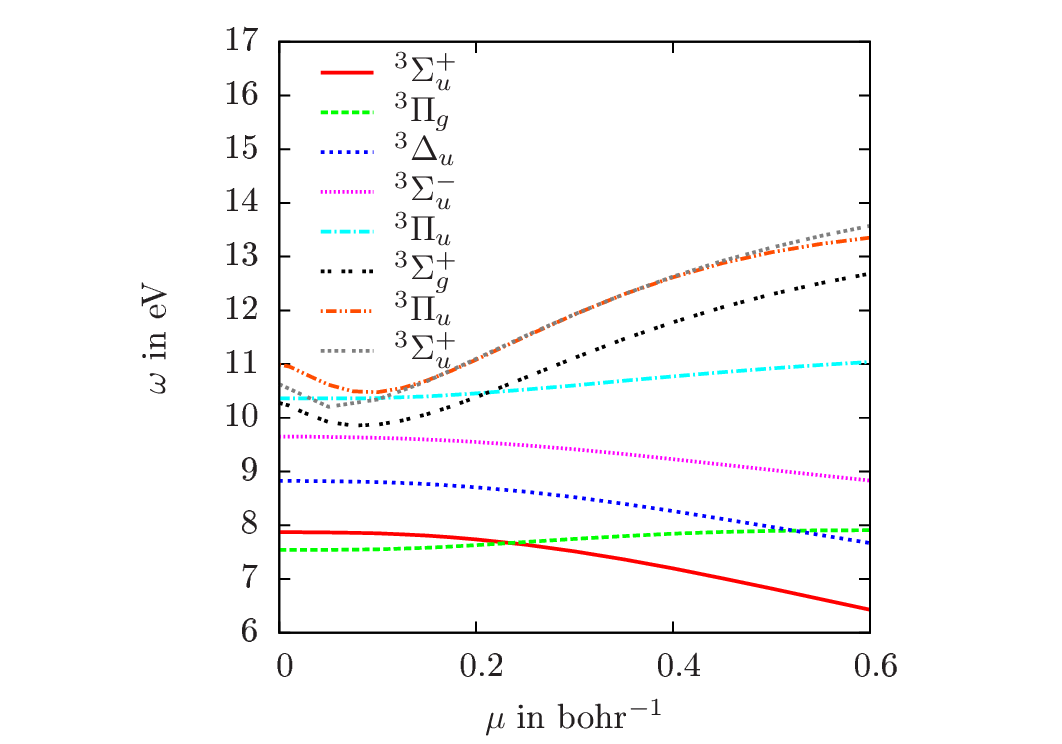}} 
\caption{Singlet (a) and triplet (b) excitation energies of N$_2$ (in eV) with respect to the range-separation parameter $\mu$ (in bohr$^{-1}$) calculated by the linear-response RSH method with the short-range LDA exchange-correlation functional at the equilibrium geometry \cite{HubHer-BOOK-79} and with the Sadlej+ basis set.\label{fig:N2_mustudy}}
\end{figure*}

The spin-adapted linear-response RSH scheme with the short-range LDA kernels has been implemented in a development version of the quantum chemistry program MOLPRO~\cite{Molproshort-PROG-12} for closed-shell systems. The implementation includes as special cases: standard TDKS with the LDA exchange-correlation functional, and time-dependent Hartree-Fock (TDHF). The implementation also includes the possibility to perform linear-response LC calculations (with the full-range LDA correlation functional). Each calculation is done in two steps: a self-consistent ground-state calculation is first performed with a chosen energy functional, and then a linear-response excited-state calculation is performed with a chosen kernel and using the previously calculated orbitals. 

For compactness, ``TD'' will be dropped in the names of the methods and ``LDA'' will also be omitted in the names as it is the only density functional used here. Therefore, ``KS'' will denote a TDKS calculation using the LDA exchange-correlation functional, ``HF'' will stand for a TDHF calculation, ``RSH'' will denote a linear-response RSH calculation using the short-range LDA exchange-correlation functional, and ``LC'' will stand for a linear-response LC calculation using the short-range LDA exchange functional and the full-range LDA correlation functional. We will call ``RSH-TDA'' a linear-response RSH calculation with the Tamm-Dancoff approximation. For all these methods, the same functional is used for the ground-state energy calculation and the linear-response calculation. In addition, we will call ``RSH-\fHx'' a linear-response RSH calculation where only the Hartree-exchange part of the RSH kernel is used (no correlation kernel) but evaluated with regular RSH orbitals (including the short-range correlation energy functional).

We calculate vertical excitation energies and oscillator strengths of five small molecules, N$_2$, CO, H$_2$CO, C$_2$H$_4$, and C$_6$H$_6$, which have already been extensively studied theoretically~\cite{CasJamCasSal-JCP-98,TawTsuYanYanHir-JCP-04,ParEll-CP-96,PeaBenHelToz-JCP-08,SerMerNebLinRoo-JCP-93,NieJorOdd-JCP-80} and experimentally~\cite{WilHicCom-JPB-79,CloRam-ARPC-83,TayWilCom-CP-82,LesMou-JCP-77}. 
In order to have unique, comparable references, equation-of-motion coupled-cluster singles doubles (EOM-CCSD) calculations were done in the same basis with the quantum chemistry program Gaussian 09~\cite{Gaussian-PROG-09}. For each molecule, we report the first 14 excited states found with the EOM-CCSD method. Following Ref.~\onlinecite{Cas-INC-95}, we define the coefficient of the (spin-orbital) single excitation $i\to a$ in the wave function of the excited state $n$ to be
\beq
c_{n,ia}=X_{n,ia}+Y_{n,ia} = \frac{1}{\sqrt{\omega_n}}\left[ \left(\b{A}-\b{B} \right)^{1/2} \b{Z}_n \right]_{ia},
\eeq
but other choices for analyzing the eigenvectors are possible, such as defining the weight of the single excitation $i\to a$ to be $w_{n,ia}=X_{n,ia}^2-Y_{n,ia}^2$~\cite{FurRap-INC-05}. Each excited state was thus assigned by looking at its symmetry and the leading orbital contributions to the excitation.
When several excited states of the same symmetry and the same leading orbital contributions were obtained, the assignment was done by increasing order of energy. Some assignments for C$_2$H$_4$ and C$_6$H$_6$ were difficult and are explained in Tables~\ref{C2H4table} and~\ref{C6H6table}.
The Sadlej basis sets~\cite{Sad-CCCC-88,SadUrb-JMST-91} were developed to describe the polarizability of valence-like states. As the description of Rydberg states requires more flexibility, they were augmented with more diffuse functions to form the Sadlej+ basis sets~\cite{CasJamCasSal-JCP-98} that we use here. The molecules are taken in their experimental geometries~\cite{HubHer-BOOK-79,LeF-MP-91,GurVeyAlc-book-89,Her-BOOK-66}. 

The C$_2$H$_4$-C$_2$F$_4$ dimer~\cite{DreWeiHea-JCP-03,MarParHea-PCCP-11,TawTsuYanYanHir-JCP-04} was studied in its cofacial configuration along the intermolecular distance coordinate $R$ in the standard 6-31G* basis set. A geometry optimization was performed during the self-consistent ground-state calculation for each method. The CT excitation was identified by assigning the molecular orbitals involved in the excitations either to C$_2$H$_4$ or C$_2$F$_4$, using the visualization program MOLDEN~\cite{Molden-PROG}.

\section{Results and discussion}
\label{sec:results}

\subsection{Variation of the RSH excitation energies with the range-separation parameter}

When the range-separation parameter is zero, $\mu=0$, the long-range HF exchange vanishes and the short-range exchange-correlation functional reduces to the usual full-range one, therefore the RSH method is equivalent to the standard KS method in this limit. With the LDA functional, linear-response KS gives good results for the low-lying valence excitation energies but underestimates the high-lying Rydberg excitation energies. This underestimation is known to be due to the incorrect exponential asymptotic decay of the LDA exchange potential~\cite{CasJamCasSal-JCP-98}. When $\mu$ increases, long-range HF exchange replaces LDA exchange and long-range LDA correlation is removed. In the limit $\mu \fl + \infty$, RSH becomes equivalent to a HF calculation, in which Rydberg excitation energies are usually better described than in LDA but valence excitation energies can be poorly described, especially the triplet ones which can be strongly underestimated and can even become imaginary due to instabilities ($\b{A}\pm\b{B}$ in Eq.~(\ref{MZ}) are no longer positive definite).

\begin{figure}[floatfix]
\centering
\hspace*{-0.5cm}
\includegraphics[scale=0.85]{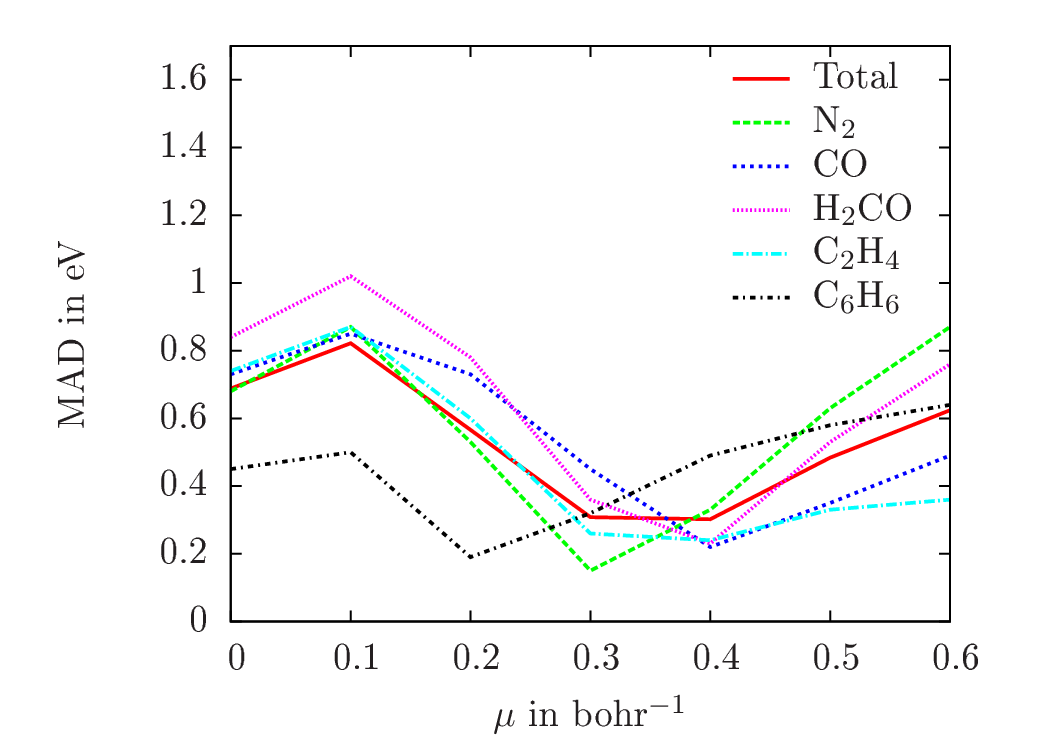}
\caption{Mean absolute deviation (MAD) in eV of the first 14 excitation energies of the N$_2$, CO, H$_2$CO, C$_2$H$_4$ and C$_6$H$_6$ molecules calculated by the linear-response RSH method with the short-range LDA exchange-correlation functional with respect to the EOM-CCSD reference as a function of the range-separation parameter $\mu$.\label{fig:MAD}}
\end{figure}

\begin{figure*}[floatfix]
\subfloat[Singlet]{\includegraphics[scale=0.85]{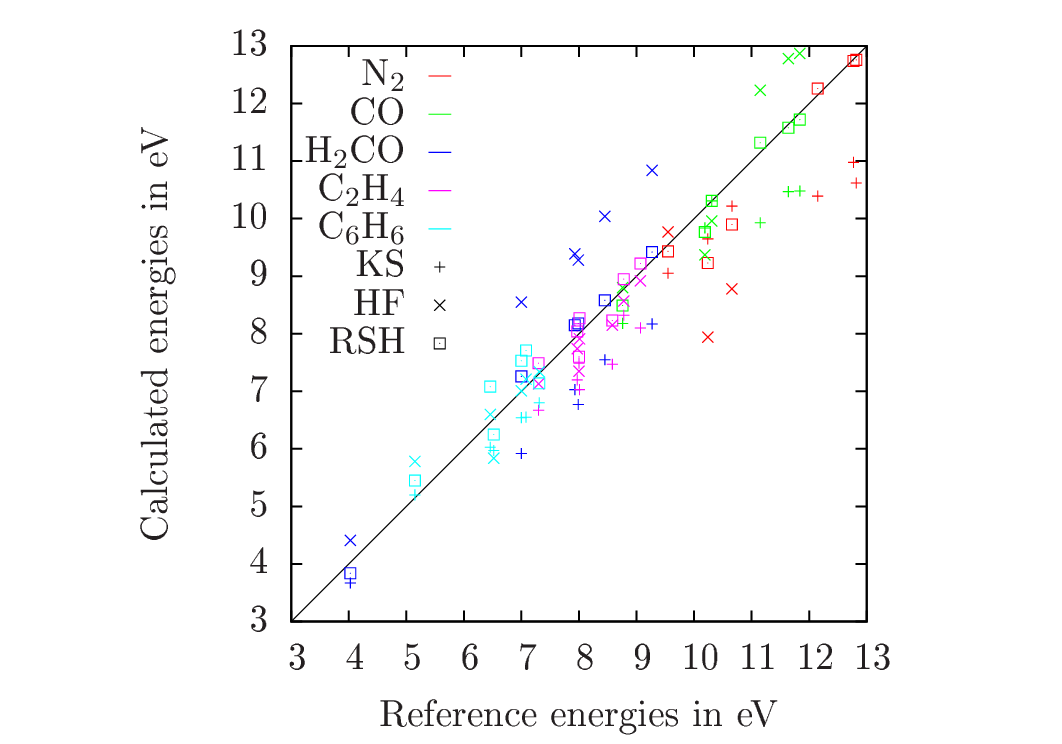}}
\hspace*{-1.5cm}
\subfloat[Triplet]{\includegraphics[scale=0.85]{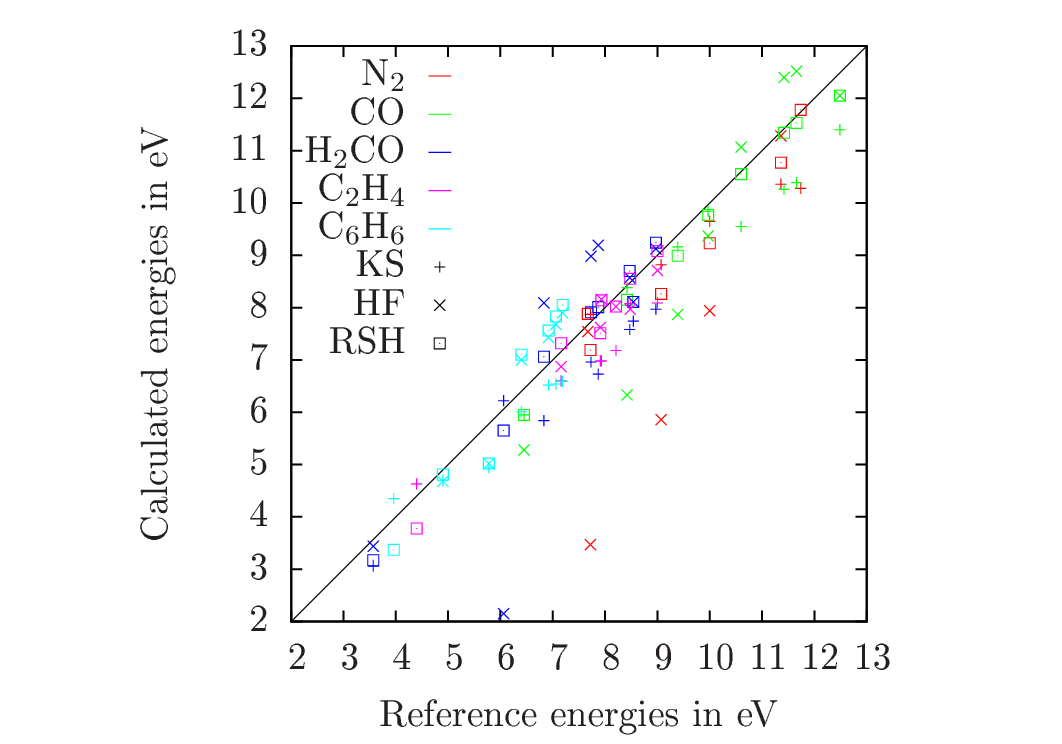}}
\caption{Singlet (a) and triplet (b) excitation energies of N$_2$, CO, H$_2$CO, C$_2$H$_4$ and C$_6$H$_6$ calculated by linear-response HF and KS (with the LDA functional), by the linear-response range-separated method RSH (with the short-range LDA functional and $\mu=0.4$ bohr$^{-1}$), as compared with the EOM-CCSD reference calculations.}
\label{fig:nuage_point}
\end{figure*}

The variation of the first few singlet and triplet RSH excitation energies of N$_2$ with respect to the range-separation parameter $\mu$ is shown in Figure~\ref{fig:N2_mustudy}. The evolution of the excitation energies is similar for both spin states, however three different trends are seen for these excitations depending on their valence or Rydberg character and their spatial symmetry. The excitation energies to the Rydberg excited states ($^1\Sigma_g^+$, $^1\Pi_u$,$^1\Sigma_u^+$, $^3\Sigma_g^+$, $^3\Sigma_u^+$, $^3\Pi_u$) which are underestimated in KS show a significant increase with $\mu$ for $\mu \gtrsim 0.1$. This behavior is quite independent of the spin and spatial symmetry of the state. The valence excited states ($^1\Pi_g$, $^1\Sigma_u^-$, $^1\Delta_u$, $^3\Sigma_u^+$, $^3\Pi_g$, $^3\Delta_u$, $^3\Sigma_u^-$,$^3\Pi_u$) which are correctly described in KS are less affected by the introduction of long-range HF exchange. However, we observe two opposite behaviors depending on the orbital character of the excitation: all the valence $\Pi$ states (corresponding to $\sigma\to\pi$ orbital transitions) have excitation energies that slowly increase with $\mu$, while for valence $\Sigma$ and $\Delta$ states (corresponding to $\pi\to\pi$ orbital transitions) the excitation energies decrease with $\mu$. As a consequence, the ordering of the states changes significantly with $\mu$. One should note that the variation of the excitation energies with $\mu$ have two causes: the variation of the orbital eigenvalues with $\mu$ in the ground-state calculation, and the variation of the kernel with $\mu$ in the linear-response calculation. Both effects can be significant.

The choice of the range-separation parameter $\mu$ is important. It has been proposed to adjust the value of $\mu$ for each system by imposing a self-consistent Koopmans' theorem condition~\cite{SteKroBae-JACS-09,SteKroBae-JCP-09}. This approach is appealing but it has the disadvantage of being non size-consistent, so we prefer to use a fixed value of $\mu$, independent of the system. In Figure~\ref{fig:MAD}, the mean absolute deviation (MAD) of the RSH excitation energies with respect to the EOM-CCSD reference is plotted as a function of $\mu$ for each molecule and for the total set. The global minimum is obtained around $\mu \approx 0.3-0.4$ bohr$^{-1}$. In all the following, we use a fixed value of $\mu = 0.4$, which is identical or similar to the value used in other range-separated TDDFT approaches~\cite{RohMarHer-JCP-09,Per-JCP-12,FroKneJen-JCP-13}. We note however that the fact that the minimum of the MAD for C$_6$H$_6$ is around $\mu=0.2$ shows that the optimal value of $\mu$ can substantially depend on the system. In particular, the presence of a triplet near-instability favors smaller values of $\mu$.

\subsection{Accuracy of the RSH excitation energies and oscillator strengths}

The excitation energies and oscillator strengths for each method and each molecule are given in Tables~\ref{N2table}-\ref{C6H6table}. Mean absolute deviations with respect to the EOM-CCSD reference are also given for the valence, the Rydberg and all the excitation energies. We also report the position of the ionization threshold for each DFT method, as given by the opposite of the HOMO orbital energy. The excitation energies for all molecules are also plotted in Figure~\ref{fig:nuage_point}. As expected, KS gives reasonably small errors for the valence excitation energies (MAD between 0.36 and 0.72 eV) but deteriorates for the Rydberg ones (MAD between 0.49 and 1.83 eV) which are largely underestimated, as seen in Figure~\ref{fig:nuage_point}. As well known~\cite{CasJamCasSal-JCP-98}, in KS with the LDA functional, the ionization energy is much too small, resulting in most of the Rydberg states and some of the valence states being in the continuum above the ionization threshold, and which are thus very much dependent on the basis set. This problem is absent in HF and range-separated approaches which correctly push up the ionization threshold. The HF excitation energies are usually larger than the reference ones except for the first triplet excitation energies which are much too small or even imaginary because of the HF triplet (near-)instability. Overall, HF gives relatively large total MADs (between 0.59 and 1.62 eV).

The RSH excitation energies are in general intermediate between KS and HF ones and in good agreement with the EOM-CCSD ones. The valence and Rydberg excitation energies are treated with a more uniform accuracy (MAD between 0.06 and 0.61 eV). However, the first triplet excitation energies are affected by the HF triplet near-instability and can be very underestimated. This effect is particularly important for the first triplet excitation energy of C$_2$H$_4$ and C$_6$H$_6$ as shown in Tables~\ref{C2H4table} and~\ref{C6H6table}. This underestimation is largely cured by the Tamm-Dancoff approximation, as shown by the RSH-TDA results. The quality of the other excitation energies is not deteriorated with this approximation, so that RSH-TDA gives overall smallest MADs than RSH. However, the oscillator strengths which were relatively good in \RSH tend to be overestimated for excitations to valence states in RSH-TDA. This has been connected with the fact that the TDA oscillator strengths violate the Thomas-Reiche-Kuhn sum rule. The present RSH results give thus very much the same trends already observed with other types of range-separated TDDFT approaches~\cite{TawTsuYanYanHir-JCP-04,PeaCohToz-PCCP-06,LivBae-PCCP-07,CuiYan-MP-10,PeaToz-JPCA-12}.

\begin{figure}[floatfix]
\hspace*{-0.5cm}
\includegraphics[scale=0.85]{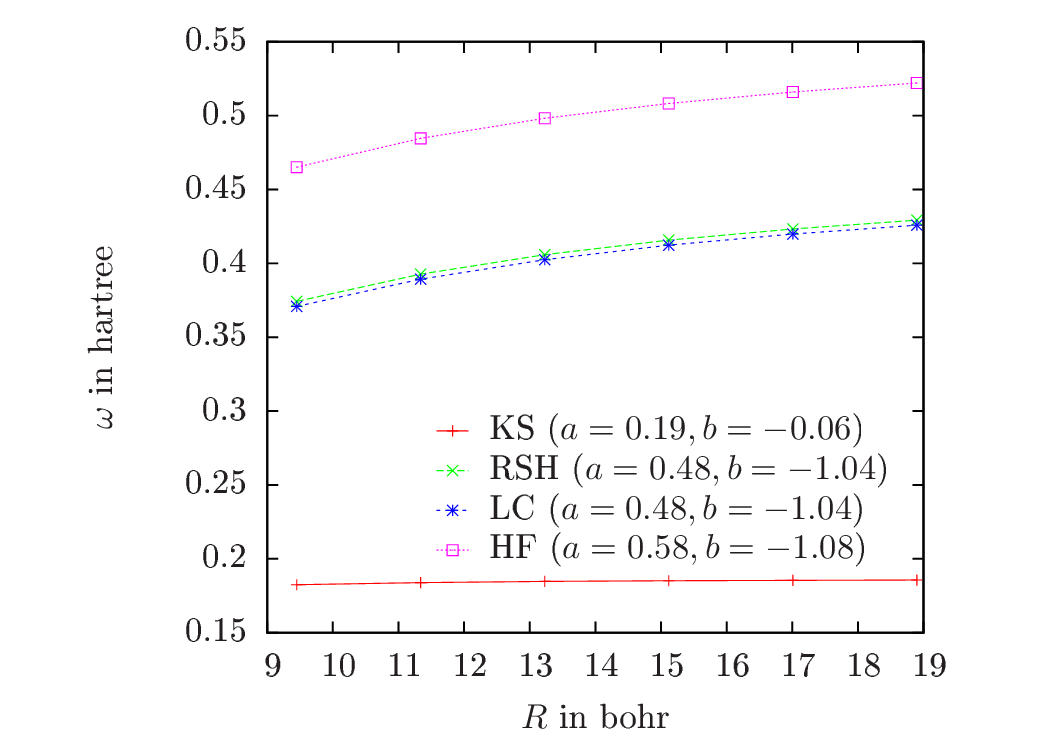}
\caption{First charge transfer excitation energy of the C$_2$H$_4$-C$_2$F$_4$ dimer calculated by linear-response HF and KS (with the LDA functional) and by the linear-response range-separated methods RSH and LC (with the short-range LDA functional and $\mu=0.4$~bohr$^{-1}$) using the 6-31G* basis. A fit of the form $a + b/R$ was performed. The fitted parameters $a$ and $b$ are given in atomic units.\label{dimer}}
\end{figure}

The first singlet CT excitation energy in the C$_2$H$_4$-C$_2$F$_4$ dimer along the intermolecular distance coordinate $R$, for $R$ between 5 and 10 {\AA} (i.e. between 9.45 and 18.90 bohr), is given in Fig.~\ref{dimer}. This excitation corresponds to an electron transfer from the HOMO of C$_2$F$_4$ to the LUMO of C$_2$H$_4$. Therefore, its energy must behave asymptotically as $I_{C_2F_4} -A_{C_2H_4} -1/R$, where $I_{C_2F_4}$ is the ionization potential of the tetrafluoroethylene and $A_{C_2H_4}$ is the electron affinity of ethylene. A fit of the form $a + b/R$ was performed and the fitted parameters are shown in Fig.~\ref{dimer}. The well-known deficiency of KS with the LDA functional to describe the $-1/R$ dependence of such excitations is observed as it gives a parameter $b$ close to zero, while HF and RSH both give the expected correct asymptotic behavior in $-1/R$ thanks to the non-locality of their exchange kernel~\cite{DreWeiHea-JCP-03}.

\subsection{Effect of the LDA correlation}

Tables~\ref{N2table}-\ref{C6H6table} also report results obtained with the \LC scheme using the short-range exchange LDA functional and the full-range LDA correlation functional. The comparison with the RSH results allows one to see the global effect of long-range LDA correlation in the ground-state calculation and in the linear-response kernel. The RSH and \LC excitation energies are globally quite close to each other, the largest difference being of 0.2 eV for the $^3 \Pi$ Rydberg state of the CO molecule. In most cases, the \LC excitation energies are slightly larger than the RSH ones. In comparison to the RSH scheme, the \LC scheme gives slightly smaller MADs (by 0.01 to 0.08 eV) for valence excitation energies, but with the exception of CO it gives larger MADs (by 0.07 to 0.09 eV) for Rydberg excitation energies. The RSH and \LC oscillator strengths are quite similar. This shows that long-range LDA correlation has a quite small effect for the systems and states considered here, and can be disregarded without much consequence.

The first CT excitation energy in the C$_2$H$_4$-C$_2$F$_4$ dimer obtained with the \LC scheme is also reported in Fig.~\ref{dimer}. Not surprisingly, the RSH and LC curves have the same $-1/R$ behavior, which is given by the long-range HF exchange kernel, and are essentially on-top on each other, showing that long-range LDA correlation has almost no effect on the HOMO and LUMO orbital energies.

To investigate the effect of the short-range LDA correlation kernel, Tables~\ref{N2table}-\ref{C6H6table} also report RSH-{\fHx} results obtained with regular RSH orbitals but no correlation kernel at all. Removing the short-range LDA correlation kernel tends to yield larger singlet excitation energies and smaller triplet excitation energies. This is not unexpected since the singlet LDA correlation kernel is negative and the triplet LDA correlation kernel is positive, as shown in Figure~\ref{fig:fc}. In comparison to the RSH results, RSH-{\fHx} gives quite similar singlet valence excitation energies and Rydberg excitation energies, but much lower triplet valence excitation energies (sometimes by as much as 0.5 eV), leading to significantly larger MADs for valence excitations. The short-range part of the LDA correlation kernel is thus important and cannot be neglected.

\section{Conclusion}
\label{sec:conclusion}

We have studied a linear-response range-separated scheme, which combines a long-range HF exchange kernel with a short-range LDA exchange-correlation kernel, for calculating electronic excitation energies and oscillator strengths of molecular systems. It is a first-level approximation before adding an explicit treatment of long-range correlation. It can also been seen as an alternative to the widely used linear-response LC scheme which combines a long-range HF exchange kernel with a short-range DFT exchange functional and a full-range DFT correlation functional.

Tests on the N$_2$, CO, H$_2$CO, C$_2$H$_4$, and C$_6$H$_6$ molecules have shown that a reasonable value for the range-separation parameter is $\mu=0.4$ bohr$^{-1}$, which is consistent with what was previously reported in the literature for other types of range-separated TDDFT methods. Just like in the LC scheme, the introduction of long-range HF exchange in the present method corrects the well-known underestimation of high-lying Rydberg excitation energies of standard TDDFT using (semi)local density-functional approximations, but also leads to underestimated excitation energies to low-lying spin-triplet valence states. This latter problem is known to be associated with the presence of HF triplet near-instabilities and is largely cured by the Tamm-Dancoff approximation which leads to a relatively uniform accuracy for all excitation energies, but possibly at the cost of deteriorating the oscillator strengths. As expected, tests on the first CT excitation energy in the C$_2$H$_4$-C$_2$F$_4$ have shown that the present range-separated TDDFT method also correctly describe this kind of excitations.

For the systems and states considered here the presence of long-range LDA correlation in the ground-state calculation and in the linear-response kernel has a quite small effect, so that the present method gives results very similar to the ones given by the LC scheme. Long-range LDA correlation can therefore be disregarded. In contrast, the short-range LDA correlation kernel is important for singlet $\to$ triplet valence excitation energies and cannot be neglected. This work thus suggests that the present range-separated TDDFT scheme is a reasonable starting approximation for describing electronic excitation energies. The next step of this work is then to add to the present method an explicit frequency-dependent long-range correlation kernel derived from perturbation theory, e.g. in the spirit of Refs.~\onlinecite{RomSanBerSotMolReiOni-JCP-09,HuiCas-ARX-10}, which would add the possibility of describing double excitations.

\section*{Acknowledgments}
It is a pleasure to dedicate this paper to Trygve Helgaker who has made outstanding contributions to many areas of quantum chemistry, including density-functional theory. We thank E. Fromager (Strasbourg, France) and K. Pernal (\L\'od\'z, Poland) for discussions.

\appendix
\section{Spin-adapted kernels} \label{appendix:spin-kernel}

For spin-restricted closed-shell calculations, spin-singlet and spin-triplet excitations can be decoupled~\cite{PetGro-IJQC-96,BauAhl-CPL-96,GisSniBae-CPC-99} (see also Refs.~\onlinecite{TouZhuSavJanAng-JCP-11,AngLiuTouJan-JCTC-11}). The non-spin-flip part of the coupling matrix $\mK$ of Eq.~(\ref{eq:Kiajb}) has the following spin block structure
\beq 
\mK =
\begin{pmatrix}
\mK_{\up, \up} & \mK_{\up,\down} \\
\mK_{\down,\up} & \mK_{\down,\down}
\end{pmatrix},
\eeq
where the matrix blocks $\mK_{\sigma,\sigma'}$ have elements of the form $\mK_{i\sigma \, a\sigma, \, j\sigma' \, b\sigma'}$ with $i$, $j$, and $a$, $b$ referring to occupied and virtual spatial orbitals, respectively. The matrix $\mK$ can be brought to a block diagonal form by rotation in spin space
\beq 
\tilde{\mK} =
\begin{pmatrix}
{^1}\mK & \b{0} \\
\b{0} & {^3}\mK
\end{pmatrix},
\eeq
with a singlet component
\beq
^1 \mK = \dfrac{\mK_{\up, \up} + \mK_{\up,\down} + \mK_{\down,\up} + \mK_{\down,\down}}{2},
\eeq
and a triplet component
\beq
^3 \mK = \dfrac{\mK_{\up, \up} - \mK_{\up,\down} - \mK_{\down,\up} + \mK_{\down,\down}}{2}.
\eeq
This directly leads to the singlet and triplet RSH coupling matrices of Eqs.~(\ref{eq:1Kiajb}) and~(\ref{eq:3Kiajb}), where the singlet and triplet short-range exchange-correlation kernels are defined as
\beq
\begin{split}
^1f_{xc}^{sr}(\mr_1,\mr_2) = &\Bigl[ f_{xc,\up \up}^{sr}(\mr_1,\mr_2)+f_{xc,\up \down}^{sr}(\mr_1,\mr_2)\\
&+f_{xc,\down \up}^{sr}(\mr_1,\mr_2)+f^{sr}_{xc,\down \down}(\mr_1,\mr_2) \Bigl]/4,
\end{split}
\label{eq:1f}
\eeq
and
\beq
\begin{split}
^3f_{xc}^{sr}(\mr_1,\mr_2) = &\Bigl[ f_{xc,\up \up}^{sr}(\mr_1,\mr_2)-f_{xc,\up \down}^{sr}(\mr_1,\mr_2)\\
&-f_{xc,\down \up}^{sr}(\mr_1,\mr_2)+f^{sr}_{xc,\down \down}(\mr_1,\mr_2) \Bigl]/4.
\end{split}
\label{eq:3f}
\eeq
The different spin components of the kernel can be expressed with the second-order functional derivatives of the corresponding energy functional $E_{xc}^{sr}[n,m]$ with respect to the density $n$ and the spin magnetization density $m$,
\begin{equation}
\begin{split}
f_{xc,\up\up}^{sr}(\mr_1,\mr_2) = \dfrac{\delta^2 E_{xc}^{sr}[n,m]}{\delta n_\up(\mr_1) \delta n_\up(\mr_2)} \;\;\;\;\;\;\;\;\;\;\;\;\;\;\;\;\;\;\;\;\;\;\;\;\;\;\;\;\;\;\;\;   \\
=\dfrac{\delta^2 E_{xc}^{sr}[n,m]}{\delta n(\mr_1) \delta n(\mr_2)}
 + 2 \dfrac{\delta^2 E_{xc}^{sr}[n,m]}{\delta n(\mr_1) \delta m(\mr_2)} + \dfrac{\delta^2 E_{xc}^{sr}[n,m]}{\delta m(\mr_1) \delta m(\mr_2)},
\end{split}
\end{equation}
and
\begin{equation}
\begin{split}
f_{xc,\up\down}^{sr}(\mr_1,\mr_2)  &= f_{xc,\down\up}^{sr}(\mr_1,\mr_2) =\dfrac{\delta^2 E_{xc}^{sr}[n,m]}{\delta n_\up(\mr_1) \delta n_\down(\mr_2)} 
\\ &= \dfrac{\delta^2 E_{xc}^{sr}[n,m]}{\delta n(\mr_1) \delta n(\mr_2)} - \dfrac{\delta^2 E_{xc}^{sr}[n,m]}{\delta m(\mr_1) \delta m(\mr_2)},
\end{split}
\end{equation}
and
\begin{equation}
\begin{split}
f_{xc,\down\down}^{sr}(\mr_1,\mr_2)  = \dfrac{\delta^2 E_{xc}^{sr}[n,m]}{\delta n_\down(\mr_1) \delta n_\down(\mr_2)} \;\;\;\;\;\;\;\;\;\;\;\;\;\;\;\;\;\;\;\;\;\;\;\;\;\;\;\;\;\;\;\;
\\ = \dfrac{\delta^2 E_{xc}^{sr}[n,m]}{\delta n(\mr_1) \delta n(\mr_2)} 
 - 2 \dfrac{\delta^2 E_{xc}^{sr}[n,m]}{\delta n(\mr_1) \delta m(\mr_2)} + \dfrac{\delta^2 E_{xc}^{sr}[n,m]}{\delta m(\mr_1) \delta m(\mr_2)}.
 \end{split}
\end{equation}
The mixed derivative with respect to $n$ and $m$ cancels out in Eqs.~(\ref{eq:1f}) and~(\ref{eq:3f}) and we finally obtain the singlet and triplet kernels of Eqs.~(\ref{eq:1fxcsr}) and~(\ref{eq:3fxcsr}).

\section{Short-range LDA exchange-correlation functional}

\subsection{Short-range LDA exchange}\label{appendix:Exsr}

The short-range spin-independent LDA exchange energy density is a function of the density $n$ (or equivalently of the Wigner-Seitz radius $r_s= (3/(4\pi n))^{1/3}$) and of the range-separation parameter $\mu$, and writes
\begin{equation}
e_{x}^{sr} = n \, (\epsilon_x - \epsilon_{x}^{lr}),
\end{equation}
where $\epsilon_x$ is the full-range exchange energy per particle of the homogeneous electron gas,
\beq
\epsilon_x = -  \dfrac{27\alpha^2}{16 \, r_s},
\eeq 
with $\alpha =  (4/(9\pi))^{1/3}$, and $\epsilon_{x}^{lr}$ is the long-range exchange energy per particle of the homogeneous electron gas \cite{Sav-INC-96,TouSavFla-IJQC-04}
\begin{equation}
\begin{split}
\epsilon_{x}^{lr} = &
-\dfrac{9\alpha^2 \, A}{r_s} 
	 \Bigl[ \sqrt{\pi} \, \mathrm{erf} \left( \dfrac{1}{2A} \right)    \\
	& + \left(2A- 4A^3 \right) e ^{-1/(4A^2)}  - 3A + 4A^3 \Bigl],
\end{split}
\end{equation}
with $A = \alpha \, \mu \, r_s/2$.

\subsubsection{Large $\mu$ behavior}

In the limit of a very short-range interaction ($\mu\to +\infty$) or in the low-density limit ($n\to0$ or $r_s\to\infty$), i.e. $A\to\infty$, the short-range exchange energy density $e_{x}^{sr}$ goes to zero with the following asymptotic expansion
\beq
e_{x}^{sr} = - \dfrac{3\,n}{16 \, r_s^3 \, \mu^2} +\dfrac{9 \, n}{320 \, \alpha^2 \, r_s^5 \, \mu^{4}} + O\left(\dfrac{1}{\mu^{6}}\right),
\eeq
and the corresponding short-range exchange kernel, i.e. the second-order derivative with respect to $n$, expands as
\beq
\dfrac{\partial^{2} e_{x}^{sr}}{\partial n^{2}} = - \dfrac{\pi}{2 \mu^2} +\dfrac{\pi}{6 \, \alpha^2 \, r_s^{2} \, \mu^{4}} + O\left(\dfrac{1}{\mu^{6}}\right).
\eeq

\subsubsection{Small $\mu$ behavior}

In the limit of the Coulombic interaction ($\mu\to0$) or in the high-density limit ($n\to +\infty$ or $r_s\to 0$), i.e. $A\to0$, the short-range exchange energy density $e_{x}^{sr}$ reduces to the full-range exchange energy density $e_{x} = n \,\epsilon_x$ with the following expansion
\beq
e_{x}^{sr} = e_x  +\dfrac{\mu \, n}{\sqrt{\pi}} - \dfrac{3 \, \alpha \, r_s \, \mu^2 \, n}{2\pi}+\dfrac{\mu^4}{6\pi^3}+  O \left( e^{-1/\mu^2} \right),
\eeq
and the short-range exchange kernel behaves as
\beq
\begin{split}
\dfrac{\partial^2 e_{x}^{sr}}{\partial n^2} = \dfrac{\partial^2 e_{x}}{\partial n^2} + \pi \, \alpha^4 \, r_s^4 \, \mu^2 +  O \left( e^{-1/\mu^2} \right),
\end{split}
\label{d2exsrdn2}
\eeq
with the full-range exchange kernel
\beq
\begin{split}
\dfrac{\partial^2 e_{x}}{\partial n^2} = -\pi \, \alpha^2 \, r_s^2.
\end{split}
\label{d2exdn2}
\eeq
Taking the ratio of Eqs.~(\ref{d2exsrdn2}) and~(\ref{d2exdn2}), it is seen the short-range exchange kernel reduces to the full-range one when
\beq
\alpha^2\mu^2 r_s^2 \ll 1,
\eeq
i.e. $r_s \ll 4.8$ for $\mu=0.4$.

\subsection{Short-range LDA correlation}\label{appendix:Ecsr}

The short-range spin-dependent LDA correlation energy density is a function of the density $n$, the spin magnetization density $m$ (or equivalently of $r_s$ and $\zeta = m/n$), and of the range-separation parameter $\mu$, and writes
\begin{equation}
e_c^{sr} = n \, (\epsilon_c - \epsilon_c^{lr}),
\end{equation}
where $\epsilon_c$ is the full-range correlation energy per particle of the homogeneous electron gas~\cite{PerWan-PRB-92}, and $\epsilon_c^{lr}$ is the correlation energy per particle of a homogeneous electron gas with the long-range electron-electron interaction, as fitted by Paziani {\it et al.}~\cite{PazMorGorBac-PRB-06} on quantum Monte Carlo calculations with imposed exact limits,
\begin{equation}
\begin{split}
\epsilon_{c}^{lr}= 
	& \Bigl[ \phi_2(\zeta)^3 Q\!\left( \dfrac{\mu \sqrt{r_s}}{\phi_2(\zeta)} \right) + a_1(r_s,\zeta) \mu^3 
	\\ & +a_2(r_s,\zeta) \mu^4 
	 + a_3(r_s,\zeta) \mu^5 + a_4(r_s,\zeta) \mu^6 
	\\ & + a_5(r_s,\zeta) \mu^8 \Bigl] \times \dfrac{1}{\left(1 + b_0^2(r_s)\mu^2 \right)^4},
\end{split}
\end{equation}
where $\phi_2(\zeta)=[(1+\zeta)^{2/3}+(1-\zeta)^{2/3}]/2$ and the other functions are given in Ref.~\onlinecite{PazMorGorBac-PRB-06}.

The derivatives of $e_c^{sr}$ with respect to $n$ and $m$ are easily expressed in terms of the derivatives of $\epsilon_c^{sr}$ with respect to $r_s$ and $\zeta$. The first-order derivatives are
\begin{equation}
\begin{split}
\dfrac{\partial e_c^{sr}}{\partial n} 
	 = & -\dfrac{r_s}{3}\dfrac{\partial \epsilon_c^{sr}}{\partial r_s} + \epsilon_c^{sr} ,\\
\dfrac{\partial e_c^{sr}}{\partial m} = & \dfrac{\partial \epsilon_c^{sr}}{\partial \zeta},
\end{split}
\end{equation}
and the second-order derivatives are
\begin{equation}
\begin{split}
\dfrac{\partial^2 e_c^{sr}}{\partial n^2} &= -\dfrac{r_s}{9n} \left( 2 \dfrac{\partial \epsilon_c^{sr}}{\partial r_s}  -r_s  \dfrac{\partial^2 \epsilon_c^{sr}}{\partial r_s^2} \right) ,\\
\dfrac{\partial^2 e_c^{sr}}{\partial m^2} &=\dfrac{1}{n}  \dfrac{\partial^2 \epsilon_c^{sr}}{\partial \zeta^2}.
\end{split}
\end{equation}
For spin-restricted closed-shell calculations, we just need to evaluate them at $\zeta=0$.

\subsubsection{Large $\mu$ behavior}

The leading terms of the asymptotic expansion for $\mu\to +\infty$ of the short-range correlation energy density $e_{c}^{sr}$ can be expressed with the on-top pair-density of the homogeneous electron gas~\cite{PazMorGorBac-PRB-06}. In the low-density limit $r_s\to +\infty$ (or the strong-interaction limit $\lambda\to +\infty$ of the adiabatic connection), it simplifies to
\beq
\begin{split}
e_c^{sr}\Bigl|_{r_s\to +\infty} =  -\dfrac{3 \left(1-\zeta^2\right) n}{16 \, r_s^3 \, \mu^2} + O\left(\dfrac{1}{\mu^4}\right).
\end{split}
\eeq
In this limit, the associated singlet and triplet short-range correlation kernels, i.e. the second-order derivatives of $e_c^{sr}$ with respect to $n$ and $m$ evaluated at $\zeta=0$, have the following expansions
\beq
\dfrac{\partial^2 e_c^{sr}}{\partial n^2}\Biggl|_{\zeta=0, \, r_s\to +\infty}=- \dfrac{\pi}{2 \mu^2} + O\left(\dfrac{1}{\mu^4}\right),
\eeq
\beq
\dfrac{\partial^2 e_c^{sr}}{\partial m^2}\Biggl|_{\zeta=0, \, r_s\to +\infty}= \dfrac{\pi}{2 \mu^2} + O\left(\dfrac{1}{\mu^4}\right).
\eeq

\subsubsection{Small $\mu$ behavior}

In the limit $\mu\to0$, the short-range correlation energy density $e_{c}^{sr}$ reduces to the full-range correlation energy density $e_{c} = n \,\epsilon_c$ with the following expansion~\cite{PazMorGorBac-PRB-06}
\beq
\begin{split}
e_c^{sr} = e_c + \dfrac{3 \, \alpha \, \phi_2(\zeta) \, r_s \, \mu^2 \, n}{2\pi}+ O(\mu^3).
\end{split}
\eeq
It can easily be shown that the singlet and triplet short-range correlation kernels, evaluated at $\zeta=0$, approach the corresponding full-range kernels with the same leading term
\beq
\dfrac{\partial^2 e_c^{sr}}{\partial n^2}\Biggl|_{\zeta=0}=\dfrac{\partial^2 e_c}{\partial n^2}\Biggl|_{\zeta=0} - \pi \, \alpha^4 \, r_s^4 \, \mu^2+ O(\mu^3),
\label{d2ecsrdn2}
\eeq
\beq
\dfrac{\partial^2 e_c^{sr}}{\partial m^2}\Biggl|_{\zeta=0}= \dfrac{\partial^2 e_c}{\partial m^2}\Biggl|_{\zeta=0} - \pi \, \alpha^4 \, r_s^4 \, \mu^2+ O(\mu^3).
\label{d2ecsrdm2}
\eeq

In the high-density limit $r_s \fl 0$, the expansion of the full-range correlation energy density has the form~\cite{PerWan-PRB-92}
\beq
e_c=n \left[ c_0(\zeta)\ln r_s-c_1(\zeta) + O\left(r_s\ln r_s\right) \right].
\eeq
The expansion of the singlet full-range correlation kernels is
\beq
\begin{split}
\dfrac{\partial^2 e_c}{\partial n^2}\Biggl|_{\zeta=0}=& -c_0(0) \, \pi^2 \, \alpha^3 \, r_s^3 + O( r_s^4\ln r_s)  ,
\end{split}
\label{d2ecdn2}
\eeq
with $c_0(0) = (1-\ln2)/\pi^2$, and the expansion of the triplet full-range correlation kernels is found from the the correlation part of the spin stiffness $\alpha_c(r_s) = (\partial^2 \epsilon_c/\partial \zeta^2)_{\zeta=0}$
\beq
\begin{split}
\dfrac{\partial^2 e_c}{\partial m^2}\Biggl|_{\zeta=0} & =3\pi^2 \, \alpha^3 \, r_s^3 \, \alpha_c(r_s) \\
& = 3\pi^2 \, \alpha^3 \, r_s^3 \left[ A_\alpha \ln r_s + C_\alpha + O(r_s \ln r_s) \right],
\end{split}
\label{d2ecdm2}
\eeq
where $A_\alpha=-1/(6\pi^2)$ and $C_\alpha=0.0354744$~\cite{VosWilNus-CJP-80}. Comparing Eqs.~(\ref{d2ecsrdn2}) and~(\ref{d2ecdn2}), it is seen the singlet short-range correlation kernel reduces to the full-range one when
\beq
\dfrac{\pi \, \alpha \, r_s \, \mu^2}{1-\ln 2} \ll 1,
\eeq
i.e.  $r_s \ll 1.2$ for $\mu=0.4$, and, comparing Eqs.~(\ref{d2ecsrdm2}) and~(\ref{d2ecdm2}),  the triplet short-range correlation kernel reduces to the full-range one when
\beq
 \dfrac{\alpha \, r_s \, \mu^2}{3\pi(A_\alpha\ln r_s + C_\alpha)} \ll 1,
\eeq
i.e.  $r_s \ll 2.4$ for $\mu=0.4$.

\begingroup
\squeezetable
\begin{table*}[floatfix]
	\centering
	\scalebox{1}{
		\begin{tabular}{|c c|c H H H c c c H H H c c H H c H|}
		\hline
State  	& Transition 								& \hspace*{0.4 cm}  KS \hspace*{0.4 cm} 	& LDAx	& KS-TDA	& TDHF/ LDA TDA& RSH	& RSH-TDA	&  \LC 	& \LC-TDA	& LC-fHx & LC-fHx-TDA 	& RSH-\fHx	& HF 	& TDA-HF	& LC-BOP 	& EOM-CCSD & Expt \\
\hline \hline
\multicolumn{18}{c}{ Valence excitation energies (eV)} \\
\hline
$^3\Sigma_u^+$	& $1\pi_u \rightarrow 1\pi_g$		& 7.87	& 7.33 & 7.87	&  & 7.19	& 7.41	& 7.31	& 7.51	& 6.65 & 6.93 	& 6.65		& 3.47  & 6.23	& 7.37	& 7.72 & 7.75	\\
$^3 \Pi_g$		& $3\sigma_g \rightarrow 1\pi_g$	& 7.54	& 7.29 & 7.54 	&  & 7.84	& 7.87	& 7.88	& 7.90	& 7.66 & 7.69 	& 7.67		& 7.62  & 7.99	& 7.79	& 8.16 & 8.04	\\
$^3\Delta_u$	& $1\pi_u \rightarrow 1\pi_g$		& 8.82	& 8.57 & 8.82	&  & 8.26	& 8.39	& 8.31 	& 8.43	& 8.02 & 8.18 	& 8.03		& 5.86  & 7.32	& 8.29	& 9.07 & 8.88	\\
$^1\Pi_g$		& $3\sigma_g \rightarrow 1\pi_g$	& 9.05	& 9.07 & 9.12	&  & 9.43	& 9.53	& 9.43	& 9.52	& 9.46 & 9.56 	& 9.47		& 9.77  & 10.01	& 9.35	& 9.55 & 9.31	\\
$^3\Sigma_u^-$	& $1\pi_u \rightarrow 1\pi_g$		& 9.65	& 9.65 & 9.65	&  & 9.23	& 9.26	& 9.22	& 9.25	& 9.22 & 9.25 	& 9.23		& 7.94  & 8.50	& 9.34	& 10.00 & 9.67	\\
$^1\Sigma_u^-$	& $1\pi_u \rightarrow 1\pi_g$		& 9.65	& 9.65 & 9.65	&  & 9.23	& 9.26	& 9.22	& 9.25	& 9.22 & 9.25 	& 9.23		& 7.94  & 8.50	& 9.34	& 10.24 & 9.92	\\
$^1\Delta_u$	& $1\pi_u \rightarrow 1\pi_g$		& 10.22	& 10.26 & 10.18	& 	& 9.90	& 9.84	& 9.90	& 9.84	& 9.94 & 9.89 	& 9.95		& 8.78  & 9.06	& 9.86	& 10.66 & 10.27	\\
$^3\Pi_u$		& $2\sigma_u \rightarrow 1\pi_g$	& 10.36	& 10.08 & 10.36	& 	& 10.77	& 10.81	& 10.82	& 10.85	& 10.52 & 10.56	& 10.53 	& 11.28 & 11.74	& 10.79	& 11.36 & 11.19 \\
\hline                                                                                                                              
\multicolumn{18}{c}{ Rydberg excitation energies (eV)} \\                                                                           
\hline                                                                                                                              
$^3\Sigma_g^+$	&$3\sigma_g \rightarrow 4\sigma_g$ 	& 10.28 & 10.22 & 10.28	& 	& 11.78 & 11.78	& 11.94 & 11.94	& 11.79 & 11.80 & 11.73 	& 13.05	& 13.12	& 11.17	& 11.74 & 12.00 \\
$^1\Sigma_g^+$	& $3\sigma_g \rightarrow 4\sigma_g$	& \textit{10.39} & 10.40 & 10.40	& 	& 12.26 & 12.29	& 12.38 & 12.42	& 12.41 & 12.44 & 12.26 	& 13.98	& 14.01	& 11.59	& 12.15 & 12.20 \\
$^3\Sigma_u^+$	& $3\sigma_g \rightarrow 3\sigma_u$	& \textit{10.62} &  &  &  & 12.63 & 12.63 & 12.87 &  &  &  & 12.59 & 14.16	& & & 12.70 &  \\
$^3\Pi_u$		& $3\sigma_g \rightarrow 2\pi_u$	& \textit{10.99} &  &  &  & 12.62 & 12.62 & 12.83 &  &  &  & 12.59 & 14.56	& & & 12.71 &  \\
$^1\Pi_u$		& $3\sigma_g \rightarrow 2\pi_u$	& \textit{10.98} & 10.99 & 10.98	& 	& 12.74 & 12.74	& 12.87	& 12.88	& 12.91 & 12.91 & 12.75 	& 13.21	& 13.23	& 12.07	& 12.77 & 12.90 \\
$^1\Sigma_u^+$	& $3\sigma_g \rightarrow 3\sigma_u$	& \textit{10.62} & 10.63 & 10.62	& 	& 12.76 & 12.77	& 12.89 & 12.91	& 12.94 & 12.96 & 12.78 	& 14.00	& 14.31	& 12.11	& 12.82 & 12.98 \\
\hline                                                                                                                              
\multicolumn{18}{c}{ Ionization threshold: $-\epsilon_{\text{HOMO}}$ (eV)} \\                                                                           
\hline                                                                                                                              
\multicolumn{2}{|c|}{                              }	& 10.38	&		&		&	& 15.34	& 15.34		& 15.76	&		&		&		&15.34			& 16.74 &		&		&		&		\\
\hline
\multicolumn{18}{c}{MAD of excitation energies with respect to EOM-CCSD (eV)} \\
\hline
Valence	& & 0.49& & & & 0.61 &  0.55 & 0.58	& & & &0.75& 1.82 & & & -&\\
Rydberg	& & 1.83& & & & 0.06 &  0.07 & 0.15	& & & &0.07& 1.35 & & & - &\\
Total	& & 1.06& & & & 0.38 &  0.34 & 0.40	& & & &0.46& 1.62 & & & -& \\
\hline                                                                                                                                                                                   
\multicolumn{18}{c}{Oscillator strengths ($\times 10^{-2}$)} \\                                                                    
\hline                                                                                                                             
$^1\Pi_u$		& $3\sigma_g \rightarrow 2\pi_u$	&  2.41	& 2.23 & 2.54	& & 9.49	& 9.54	& 12.77	& 12.73	& 11.90 & 11.78 & 9.00 		& 8.42	& 8.56  & 11.90	& 8.51 & 24.3\\
$^1\Sigma_u^+$	& $3\sigma_g \rightarrow 3\sigma_u$	&  1.06	& 0.93 & 1.16	& & 21.11	& 20.09	& 27.59	& 25.76	& 26.00 & 24.20 & 19.94		& 73.31	& 57.38	& 25.90	& 17.36 & 27.9	\\
\hline

		\end{tabular}
	}
\caption{Excitation energies and oscillator strengths of N$_2$ calculated by linear-response HF and KS (with the LDA functional), by the linear-response range-separated methods RSH, RSH-TDA, LC, and RSH-{\fHx} (with the short-range LDA functional and $\mu=0.4$ bohr$^{-1}$), and by EOM-CCSD taken as reference, using the Sadlej+ basis set. Excitation energies above the ionization threshold are indicated in italics.}
\label{N2table}
\end{table*}
\endgroup

\begingroup
\squeezetable
\begin{table*}[floatfix]
\scalebox{1}{
\begin{tabular}{|ll|c H H c c c H H H c c H H c H|}
\hline
State 	 			& Transition					& \hspace*{0.4cm} KS \hspace*{0.4cm}	& LDA-\fHx & TDA-LDA	& RSH & RSH-TDA	& \LC & LC-TDA	& LC-\fHx & LC-\fHx-TDA & RSH-\fHx & HF & HF-TDA	& LC-BOP	& EOM-CCSD &Expt \\
\hline \hline
\multicolumn{17}{c}{ Valence excitation energies (eV)} \\
\hline
$^3\Pi$ 		& $5a_1(\sigma) \rightarrow 2e_1(\pi^*)$	& 5.95			& 5.55 & 5.95	& 5.95	& 6.02	& 6.06 	& 6.12	& 5.62 & 5.71 	& 5.62	& 5.28	& 5.85	& 6.07	& 6.45 	& 6.32 \\
$^3\Sigma^+$ 	& $1e_1(\pi) \rightarrow 2e_1(\pi^*)$		& 8.38			& 7.91 & 8.38	& 8.15	& 8.28	& 8.22	& 8.35	& 7.70 & 7.88 	& 7.72	& 6.33	& 7.79	& 8.20	& 8.42 	& 8.51 \\
$^1\Pi$ 		& $5a_1(\sigma) \rightarrow 2e_1(\pi^*)$	& 8.18			& 8.23 & 8.35	& 8.49	& 8.67	& 8.49 	& 8.67	& 8.53 & 8.73 	& 8.55	& 8.80	& 9.08	& 8.44	& 8.76 	& 8.51 \\
$^3\Delta$		& $1e_1(\pi) \rightarrow 2e_1(\pi^*)$		& \textit{9.16}	& 8.95 & 9.16	& 8.99	& 9.07	& 9.02	& 9.10	& 8.78 & 8.88 	& 8.80	& 7.87	& 8.74	& 8.93	& 9.39 	& 9.36 \\
$^3\Sigma^-$	& $1e_1(\pi) \rightarrow 2e_1(\pi^*)$		& \textit{9.84}	& 9.84 & 9.84	& 9.77	& 9.79	& 9.75	& 9.77	& 9.75 & 9.77 	& 9.77	& 9.37	& 9.75	& 9.77	& 9.97 	& 9.88 \\
$^1\Sigma^-$	& $1e_1(\pi) \rightarrow 2e_1(\pi^*)$		& \textit{9.84}	& 9.84 & 9.84	& 9.77	& 9.79	& 9.75 	& 9.77	& 9.75 & 9.77 	& 9.77	& 9.37	& 9.73	& 9.77	& 10.19 & 9.88 \\
$^1\Delta$		& $1e_1(\pi) \rightarrow 2e_1(\pi^*)$		& \textit{10.31}& 10.34 & 10.25	& 10.31	& 10.27	& 10.29 & 10.26	& 10.33 & 10.29 & 10.35 & 9.96	& 10.15	& 10.20	& 10.31 & 10.23 \\
$^3\Pi$			& $4a_1(\sigma) \rightarrow 2e_1(\pi^*)$	& \textit{11.40}&		&		& 12.05 & 12.07 & 12.05 &		&		&		& 11.91 & 13.05 &	&	& 12.49	& \\
\hline
\multicolumn{17}{c}{Rydberg excitation energies (eV)} \\
\hline
$^3\Sigma^+$	& $5a_1(\sigma) \rightarrow 6a_1(\sigma)$	& \textit{9.55}	& 9.40 & 9.55	& 10.55	& 10.56	& 10.72	& 10.72	& 10.48 & 10.50 &10.46 	& 11.07	& 11.18	& 9.82	& 10.60 & 10.40 \\
$^1\Sigma^+$	& $5a_1(\sigma) \rightarrow 6a_1(\sigma)$	& \textit{9.93}	& 9.95 & 9.94	& 11.32	& 11.37	& 11.36 & 11.41	& 11.41 & 11.46 &11.34 	& 12.23	& 12.27	& 10.30	& 11.15 & 10.78 \\
$^3\Sigma^+$	& $5a_1(\sigma) \rightarrow 7a_1(\sigma)$	& \textit{10.26}& 10.08 & 10.26	& 11.34	& 11.35	& 11.51	& 11.51	& 11.33 & 11.34 &11.29 	& 12.40	& 12.42	& 10.65	& 11.42 & 11.30 \\
$^1\Sigma^+$	& $5a_1(\sigma) \rightarrow 7a_1(\sigma)$	& \textit{10.47}& 10.54 & 10.49	& 11.58	& 11.59	& 11.63 & 11.64	& 11.69 & 11.69 &11.60 	& 12.78	& 12.79	& 10.76	& 11.64 & 11.40 \\
$^3\Pi$			& $5a_1(\sigma) \rightarrow 3e_1(\pi)$		& \textit{10.39}& 10.28 & 10.39	& 11.53	& 11.54	& 11.73	& 11.73	& 11.56 & 11.60 &11.49 	& 12.52	& 12.60	& 10.74	& 11.66	& 11.55 \\
$^1\Pi$ 		& $5a_1(\sigma) \rightarrow 3e_1(\pi)$		& \textit{10.48}& 10.51 & 10.49	& 11.72	& 11.72	& 11.81	& 11.82	& 11.85 & 11.86 &11.73 	& 12.87	& 12.88	& 10.86	& 11.84 & 11.53 \\

\hline                                                                                                                              
\multicolumn{17}{c}{ Ionization threshold: $-\epsilon_{\text{HOMO}}$ (eV)} \\                                                                           
\hline                                                                                                                              
\multicolumn{2}{|c|}{                              }			& 9.12	&		&		& 13.83	&13.83		& 14.27	&		&		&		&	13.83		& 15.11 &		&		&		&		\\
\hline
\multicolumn{17}{c}{ MAD of excitation energies with respect to the EOM-CCSD calculation (eV)} \\
\hline
Valence				&& 0.36 & & & 0.31 & 0.25 & 0.29	& & & & 0.45 & 0.89	& & & - &\\
Rydberg				&& 1.21 & & & 0.10 & 0.10 & 0.09	& & & & 0.13 & 0.92 & & & - & \\
Total				&& 0.73 & & & 0.22 & 0.19 & 0.20	& & & & 0.31 & 0.91	& & & - &\\
\hline
\multicolumn{17}{c}{Oscillator strengths ($\times 10^{-2}$)} \\
\hline
$^1\Pi$ 			&$5a_1(\sigma) \rightarrow 2e_1(\pi^*)$	& 8.69		& 8.50  & 13.27	& 8.64	& 12.96	& 8.73 	& 13.04	& 8.56 	& 12.72	& 8.49	& 8.55	& 11.48	& 19.92	& 8.66 	& 17.6 \\
$^1\Sigma^+$		&$5a_1(\sigma) \rightarrow 6a_1(\sigma)$& 1.84		& 1.83 	& 2.40	& 4.26	& 4.44	& 3.70  & 3.67	& 2.66 	& 2.35 	& 3.85	& 10.58	& 10.49	& 2.45	& 0.58 	& - \\
$^1\Sigma^+$		&$5a_1(\sigma) \rightarrow 7a_1(\sigma)$& 12.53		& 12.16 & 13.14	& 13.73	& 14.67	& 15.86 & 17.13	& 16.55 & 18.06 & 13.73	& 9.39	& 10.21	& 15.13	& 20.71 & - \\
$^1\Pi$ 			&$5a_1(\sigma) \rightarrow 3e_1(\pi)$	& 2.71		& 2.58 	& 2.22	& 4.72	& 4.36	& 5.45	& 5.10	& 5.22 	& 4.88 	& 4.58	& 5.13	& 4.93	& 4.07	& 4.94 	& - \\
\hline
\end{tabular}
}
\caption{Excitation energies and oscillator strengths of CO calculated by linear-response HF and KS (with the LDA functional), by the linear-response range-separated methods RSH, RSH-TDA, LC, and RSH-{\fHx} (with the short-range LDA functional and $\mu=0.4$ bohr$^{-1}$), and by EOM-CCSD taken as reference, using the Sadlej+ basis set. Excitation energies above the ionization threshold are indicated in italics.}
\label{COtable}
\end{table*}
\endgroup

\begingroup
\squeezetable
\begin{table*}[floatfix]
\centering
\scalebox{1}{
\begin{tabular}{|ll|c H H c c c H H H c c H H c H|}
\hline
State 		& Transition	& \hspace*{0.4cm} KS \hspace*{0.4cm}	& LDA-fHx & TDA-LDA	& RSH & RSH-TDA	& LC & LC-TDA	& LC-LDAx & LC-LDAx-TDA & RSH-\fHx & HF & HF-TDA	& LC-BOP	& EOM-CCSD & Expt \\
\hline \hline
\multicolumn{17}{c}{ Valence excitation energies (eV)} \\
\hline
$^3A_2$		& $2b_2(n) \rightarrow 2b_1(\pi^*)$		& 3.06			& 2.96 	& 3.06	& 3.17	& 3.19	& 3.16	& 3.20	& 3.06 	& 3.09 	& 3.08	& 3.44 	& 3.76	& 3.15	& 3.56 & 3.50 \\
$^1A_2$		& $2b_2(n)\rightarrow 2b_1(\pi^*)$		& 3.67			& 3.69 	& 3.67	& 3.84	& 3.85	& 3.82	& 3.84	& 3.84 	& 3.85 	& 3.86	& 4.41 	& 4.58	& 3.82	& 4.03 & 3.94 \\
$^3A_1$		& $1b_1(\pi)\rightarrow 2b_1(\pi^*)$	& 6.22			& 5.81 	& 6.22	& 5.65	& 5.87	& 5.74	& 5.95	& 5.25 	& 5.53 	& 5.25	& 2.15 	& 4.96	& 5.82	& 6.06 & 5.53 \\
$^3B_1$		& $5a_1(\sigma)\rightarrow 2b_1(\pi^*)$	& \textit{7.74} &		&		& 8.11	& 8.14	& 8.11	&		&		&		& 7.99 	& 8.19	&		&		& 8.54 &	\\	
\hline
\multicolumn{17}{c}{Rydberg excitation energies (eV)} \\
\hline
$^3B_2$		& $2b_2(n)\rightarrow 6a_1(\sigma)$		& 5.84			& 5.71 & 5.84	& 7.06	& 7.07	& 7.17	& 7.17	& 7.03 	& 7.04 	& 7.01	& 8.09 & 8.17	& 6.60	& 6.83 & 6.83 \\
$^1B_2$		& $2b_2(n)\rightarrow 6a_1(\sigma)$		& 5.92			& 5.95 & 5.92	& 7.26	& 7.26	& 7.30	& 7.30	& 7.33 	& 7.33 	& 7.28	& 8.55 & 8.56	& 6.74	& 7.00 & 7.09 \\
$^3B_2$		& $2b_2(n)\rightarrow 7a_1(\sigma)$		& \textit{6.96}	& 6.86 & 6.95	& 7.91	& 7.92	& 7.99	& 7.99	& 7.86 	& 7.87 	& 7.86	& 8.98 & 9.04	& 7.32	& 7.73 & 7.96 \\
$^3A_1$		& $2b_2(n) \rightarrow 3b_2(\sigma)$	& \textit{6.73}	& 6.61 & 6.72	& 8.01	& 8.01	& 8.17	& 8.17	& 8.03 	& 8.04 	& 7.96	& 9.19	& 9.24	& 7.47	& 7.87 & 7.79 \\
$^1B_2$		& $2b_2(n)\rightarrow 7a_1(\sigma)$		& \textit{7.04}	& 7.06 & 7.03	& 8.15	& 8.16	& 8.17	& 8.18	& 8.20 	& 8.21 	& 8.17	& 9.39	& 9.41	& 7.48	& 7.93 & 8.12 \\
$^1A_1$		& $2b_2(n) \rightarrow 3b_2(\sigma)$	& \textit{6.77}	& 6.80 & 6.77	& 8.18	& 8.19	& 8.27	& 8.27	& 8.30 	& 8.31 	& 8.19	& 9.28	& 9.53	& 7.57	& 7.99 & 7.97 \\
$^1A_2$		& $2b_2(n) \rightarrow 3b_1(\pi)$		& \textit{7.55}	& 7.57 & 7.55	& 8.58	& 8.58	& 8.67	& 8.67	& 8.68 	& 8.68 	& 8.58	& 10.04	& 10.04	& 7.72	& 8.45 & 8.38 \\
$^3A_2$		& $2b_2(n) \rightarrow 3b_1(\pi)$		& \textit{7.58}	&		&		& 8.57	& 8.57	& 8.70	&		&		&		& 8.56 	& 9.84	&	&	& 8.47	&	\\
$^3B_2$		& $2b_2(n)\rightarrow 8a_1(\sigma)$		& \textit{7.97} &		&		& 9.12	& 9.14	& 9.24	&		&		&		& 9.06 	& 10.24 &	&	& 8.97	&	\\	
$^1B_2$		& $2b_2(n)\rightarrow 8a_1(\sigma)$		& \textit{8.17}	& 8.20 & 8.18	& 9.42	& 9.43	& 9.49	& 9.50	& 9.51 & 9.53 	& 9.44	& 10.84	& 10.87	& 8.70	& 9.27 & 9.22 \\
\hline                                                                                                                              
\multicolumn{17}{c}{ Ionization threshold: $-\epsilon_{\text{HOMO}}$ (eV)} \\                                                                           
\hline                                                                                                                              
\multicolumn{2}{|c|}{                              }			& 6.30	&		&		& 10.63	&10.63		& 11.06	&		&		&		&	10.63		& 12.04 &		&		&		&		\\
\hline
\multicolumn{17}{c}{ MAD of excitation energies with respect to EOM-CCSD (eV)} \\             
\hline                                   
Valence		& 			& 0.46 &	& 	&  0.36	& 0.28	& 0.34 & & & & 0.50	&1.19& & &	-	&\\
Rydberg		& 			& 1.00 &	& 	&  0.18	& 0.18	& 0.26 & & & & 0.16	&1.39& & &	-	&\\
Total		& 			& 0.84 &	& 	&  0.23	& 0.21	& 0.29 & & & & 0.26	&1.34& & &	-	&\\
\hline                                   
\multicolumn{17}{c}{Oscillator strengths ($\times 10^{-2}$)} \\
\hline
$^1B_2$		& $2b_2(n)\rightarrow 6a_1(\sigma)$		& 3.13	& 3.01 	& 3.80	& 1.77	& 2.09	& 1.88	& 2.22	& 1.77 	& 2.08 	& 1.69	& 2.99	& 3.17	& 0.08	& 2.15 	& 2.8,3.2,3.8,4.13 \\
$^1B_2$		& $2b_2(n)\rightarrow 7a_1(\sigma)$		& 2.05	& 2.02 	& 2.32	& 4.58	& 5.54	& 5.04	& 5.58	& 4.95 	& 5.44 	& 4.47	& 4.46	& 4.62	& 7.17	& 4.12 	& 1.7,1.9,1.7,2.81 \\
$^1A_1$		& $2b_2(n) \rightarrow 3b_2(\sigma)$	& 4.34	& 4.21 	& 4.90	& 5.35	& 6.10	& 6.07	& 6.93	& 5.84	& 6.67 	& 5.18	& 21.31 & 16.49	& 2.21 	& 5.70 	& 3.2,3.6,3.8,6.05 \\
$^1B_2$		& $2b_2(n)\rightarrow 8a_1(\sigma)$		& 4.27	&		&		& 4.01	& 4.84	& 4.08	& 		&		&		& 3.88	& 6.65  &		&		& 4.22	&	\\
\hline
\end{tabular}
}
\caption{Excitation energies and oscillator strengths of H$_2$CO calculated by linear-response HF and KS (with the LDA functional), by the linear-response range-separated methods RSH, RSH-TDA, LC, and RSH-{\fHx} (with the short-range LDA functional and $\mu=0.4$ bohr$^{-1}$), and by EOM-CCSD taken as reference, using the Sadlej+ basis set. The molecule is oriented in the $yz$ plane along the $z$ axis. Excitation energies above the ionization threshold are indicated in italics.}
\label{H2COtable}
\end{table*}
\endgroup

\begingroup
\squeezetable
\begin{table*}[floatfix]
\centering
\scalebox{1}{
\begin{tabular}{|l l|c H H c c c H H H c c H H c H|}
\hline
%
State 		& Transition 			& \hspace*{0.4cm} KS\hspace*{0.4cm}	& KS-fHx	& KS-TDA	& RSH 	& RSH-TDA	& \LC 	& \LC-TDA	& LC-LDAx 	& LC-LDAx-TDA & RSH-\fHx	& HF & HF-TDA	& LC-BOP	& EOM-CCSD & Expt \\
\hline \hline
\multicolumn{17}{c}{ Valence excitation energies (eV)} \\
\hline
$^3B_{1u}$ 	& $1b_{3u}(\pi) \rightarrow 1b_{2g} (\pi^*)$ 	& 4.63 			& 4.18	& 4.63 	& 3.78 	& 4.13 	& 3.92 	& 4.23 	& 3.32	& 3.78	& 3.31	& 0.16 	& 3.54 	& 4.18 	& 4.41 	& 4.36 \\ 
$^1B_{1u}$ 	& $1b_{3u}(\pi) \rightarrow 1b_{2g} (\pi^*)$ 	& \textit{7.49}	& 7.50	& 7.93 	& 7.60 	& 8.03 	& 7.60 	& 8.04 	& 7.64	& 8.10 	& 7.65	& 7.35 	& 7.70	& 7.56 	& 8.00 	& 8.00 \\ 
$^3B_{1g}$ 	& $1b_{3g}(\sigma) \rightarrow 1b_{2g}(\pi^*)$			& \textit{7.18}	&		&		& 8.03	& 8.04	& 8.15	&		&		&		& 8.02	& 8.36	&		&		& 8.21	&		\\
$^1B_{1g}$	& $1b_{3g}(\sigma) \rightarrow 1b_{2g}(\pi^*)$ $^{(a)}$	& \textit{7.47}	&		&		& 8.15	& 8.16	& 8.23	&		&		&		& 8.16	& 9.36	&		&		& 8.58	&		\\ 
\hline
\multicolumn{17}{c}{Rydberg excitation energies (eV)} \\
\hline
$^3B_{3u}$ 	& $1b_{3u}(\pi) \rightarrow 4a_{1g}(\sigma)$	& 6.60 			& 6.50	& 6.60	& 7.32 	& 7.33 	& 7.44 	& 7.45 	& 7.34	& 7.34	& 7.29	& 6.87 	& 6.91 	& 7.01 	& 7.16 	& 6.98 \\ 
$^1B_{3u}$ 	& $1b_{3u}(\pi) \rightarrow 4a_{1g}(\sigma)$	& 6.67			& 6.67	& 6.66 	& 7.49 	& 7.49	& 7.55 	& 7.55 	& 7.57	& 7.57 	& 7.50	& 7.13 	& 7.14 	& 7.13 	& 7.30 	& 7.11 \\ 
$^3B_{1g}$ 	& $1b_{3u}(\pi) \rightarrow 2b_{2u}(\sigma)$	& \textit{6.98} & 6.87	& 6.98	& 7.51 	& 7.53 	& 7.50 	& 7.52 	& 7.39	& 7.41 	& 7.43	& 7.63 	& 7.66 	& 7.43 	& 7.91	& 7.79 \\
$^3B_{2g}$ 	& $1b_{3u}(\pi) \rightarrow 3b_{1u}(\sigma)$	& \textit{6.98}	&		&		& 8.16	& 8.16	& 8.26	&		&		&		& 8.14	& 7.75	&		&		& 7.93	&		\\  
$^1B_{1g}$ 	& $1b_{3u}(\pi) \rightarrow 2b_{2u}(\sigma)$ $^{(b)}$	& \textit{7.20}	& 7.20	& 7.20	& 8.04 	& 8.05 	& 8.02 	& 8.04 	& 8.04	& 8.06 	& 8.05	& 7.74	& 7.75	& 7.74 	& 7.97 	& 7.80 \\
$^1B_{2g}$ 	& $1b_{3u}(\pi) \rightarrow 3b_{1u} (\sigma)$	& \textit{7.03}	& 7.16	& 7.03 	& 8.27	& 8.27 	& 8.36 	& 8.36 	& 8.39	& 8.40 	& 8.27	& 7.91 	& 7.92	& 7.86 	& 8.01 	& 7.90 \\
$^3A_g$ 	& $1b_{3u}(\pi) \rightarrow 2b_{3u}(\pi)$		& \textit{8.08} & 7.84	& 8.08 	& 8.55  & 8.55 	& 8.77 	& 8.78 	& 8.57	& 8.58 	& 8.49	& 7.97 	& 8.01	& 7.92 	& 8.48 	& 8.15 \\
$^1A_g$ 	& $1b_{3u}(\pi) \rightarrow 2b_{3u}(\pi)$		& \textit{8.32} & 8.32	& 8.34 	& 8.95 	& 8.98 	& 9.04 	& 9.07 	& 9.07	& 9.10	& 8.97	& 8.57 	& 8.61 	& 8.10 	& 8.78	& 8.28 \\ 
$^3B_{3u}$  & $1b_{3u}(\pi) \rightarrow 5a_{1g}(\sigma)$	& \textit{8.09} & 8.10	& 8.09 	& 9.08 	& 9.09 	& 9.18 	& 9.19 	& 9.07	& 9.08	& 9.03	& 8.71 	& 8.75 	& 8.51 	& 9.00 	& 8.57 \\ 
$^1B_{3u}$  & $1b_{3u}(\pi) \rightarrow 5a_{1g}(\sigma)$	& \textit{8.10} & 8.29	& 8.10 	& 9.22 	& 9.23	& 9.26 	& 9.27 	& 9.32	& 9.32	& 9.23	& 8.92 	& 8.92 	& 8.56 	& 9.07 	& 8.62 \\ 
\hline                                                                                                                              
\multicolumn{17}{c}{ Ionization threshold: $-\epsilon_{\text{HOMO}}$ (eV)} \\                                                                           
\hline                                                                                                                              
\multicolumn{2}{|c|}{                              }			& 6.89	&		&		& 10.61	&10.61		& 11.07	&		&		&		&	10.61		& 10.23 &		&		&		&		\\
\hline
\multicolumn{17}{c}{ MAD of excitation energies with respect to EOM-CCSD (eV)} \\ 
\hline
Valence	 	&						& 0.72	& 		& 		& 0.41	& 0.23	& 0.33	&	&	&	& 0.52	& 1.46	&	&	& -	& \\ 
Rydberg		& 						& 0.76	& 		& 		& 0.18	& 0.18	& 0.26	&	& 	&	& 0.18	& 0.24	&	&	& -	& \\ 
Total		& 						& 0.74	& 		& 		& 0.24	& 0.20	& 0.28	&	&	&	& 0.27	& 0.59	&	&	& -	& \\ 
\hline
\multicolumn{17}{c}{Oscillator strengths ($\times 10^{-2}$)} \\
\hline
$^1B_{1u}$ 	& $1b_{3u}(\pi) \rightarrow 1b_{2g} (\pi^*)$	& 30.64	& 29.94	& 46.52 & 35.42 & 53.86	& 35.77 & 54.76 & 35.42	& 35.41	& 35.10	& 39.99	& 50.54 & 75.43 & 36.29 & 29 \\ 
$^1B_{3u}$ 	& $1b_{3u}(\pi) \rightarrow 4a_{1g}(\sigma)$	& 6.30 	& 6.34	& 6.90 	& 7.64	& 8.02 	& 8.14 	& 8.57 	& 7.78	& 7.78	& 7.38	&  9.08	& 9.62 	& 13.05 & 8.16 	& 4 \\ 
$^1B_{3u}$ 	& $1b_{3u}(\pi) \rightarrow 5a_{1g}(\sigma)$	& 0.02 	& 0.07	& 0.03 	& 1.26	& 1.30 	& 1.56 	& 1.60 	& 1.80	& 1.80	& 1.14	& 0.63	& 0.62 	& 0.42	& 0.61 	& - \\ 
\hline 
\end{tabular}
}
\caption{Excitation energies and oscillator strengths of C$_2$H$_4$ calculated by linear-response HF and KS (with the LDA functional), by the linear-response range-separated methods RSH, RSH-TDA, LC, and RSH-{\fHx} (with the short-range LDA functional and $\mu=0.4$ bohr$^{-1}$), and by EOM-CCSD taken as reference, using the Sadlej+ basis set. The molecule is oriented in the $yz$ plane along the $z$ axis. Excitation energies above the ionization threshold are indicated in italics.
 $^{(a)}$ and $^{(b)}$: These two excitations mix heavily in LDA~\cite{CasSal-JCP-00,HamCasSal-JCP-02} and the leading orbital contribution to the excitation changes with the range-separation parameter. Adiabatic curves respect to $\mu$ were followed to do the assignment, with state (b) defined as the lowest $^1B_{1g}$ state with orbital transitions $1b_{3g}(\sigma) \rightarrow 1b_{2g}(\pi^*)$ and $1b_{3u}(\pi) \rightarrow 2b_{2u}(\sigma)$, and state (a) defined as the second lowest one.}
\label{C2H4table}
\end{table*}
\endgroup

\begingroup
\squeezetable
\begin{table*}[floatfix]
\centering
\scalebox{1}{
\begin{tabular}{|c l|c H H c c c H H H c c H H c H|}
\hline
State 			& Transition				&\hspace*{0.4cm} KS \hspace*{0.4cm}	& KS-fHx&KS-TDA	& RSH	& RSH-TDA & \LC & \LC-TDA& LC-LDAx& LC-LDAx-TDA & RSH-\fHx & HF & HF-TDA	& LC-BOP	& EOM-CCSD & Expt \\
\hline
\hline
 \multicolumn{17}{c}{ Valence excitation energies (eV)} \\
\hline
$ ^3B_{1u}$ 	& $1e_{1g}(\pi) \rightarrow 1e_{2u}(\pi^*)$		& 4.35 	& 4.02 	& 4.35 	& 3.37 	& 3.94 	& 3.49 	& 4.02 	& 2.90 	& 3.64	& 2.88	& - 	& 3.35 	& 3.73 	& 3.96 	& 3.94 \\  
$ ^3E_{1u}$ 	& $1e_{1g}(\pi) \rightarrow 1e_{2u}(\pi^*)$ 	& 4.69 	& 4.55 	& 4.69 	& 4.81 	& 4.83 	& 4.84 	& 4.86 	& 4.69 	& 4.72 	& 4.69	& 4.68 	& 4.86 	& 4.80 	& 4.90 	& 4.76 \\ 
$ ^1B_{2u}$		& $1e_{1g}(\pi) \rightarrow 1e_{2u}(\pi^*)$ 	& 5.20 	& 5.21 	& 5.22 	& 5.45 	& 5.55 	& 5.45 	& 5.55 	& 5.47 	& 5.57 	& 5.47	& 5.78 	& 5.98 	& 5.39 	& 5.15 	& 4.90 \\ 
$ ^3B_{2u}$ 	& $1e_{1g}(\pi) \rightarrow 1e_{2u}(\pi^*)$ 	& 4.94 	& 4.88 	& 4.94 	& 5.02 	& 5.19 	& 5.05 	& 5.21 	& 4.95 	& 5.14 	& 4.95	& 5.02 	& 5.44 	& 5.03 	& 5.78	& 5.60 \\  
$ ^1B_{1u}$ 	& $1e_{1g}(\pi) \rightarrow 1e_{2u}(\pi^*)$ 	& 5.97 	& 5.99 	& 6.20 	& 6.25 	& 6.44 	& 6.24 	& 6.44 	& 6.28 	& 6.48 	& 6.29	& 5.84 	& 6.16 	& 6.21 	& 6.52 	& 6.20 \\  
$ ^1E_{1u}$ 	& $1e_{1g}(\pi) \rightarrow 1e_{2u}(\pi^*)$ 	& \textit{6.80} 	& 6.82 	& 7.18 	& 7.14 	& 7.64 	& 7.13 	& 7.63 	& 7.15 	& 7.67 	& 7.16	& 7.34 	& 7.65 	& 6.80 	& 7.30 	& 6.94 \\ 
\hline
\multicolumn{17}{c}{Rydberg excitation energies (eV)} \\
\hline
$ ^3E_{1g}$ 	& $1e_{1g}(\pi) \rightarrow 4a_{1g}(\sigma)$& 6.01	&		&		& 7.00	& 7.00	& 7.10	&		&		&		& 7.00	& 6.46	 &		&		& 6.40 &	\\
$ ^1E_{1g}$ 	& $1e_{1g}(\pi) \rightarrow 4a_{1g}(\sigma)$& 6.03 	& 6.04 	& 6.02 	& 7.08 	& 7.08 	& 7.14 	& 7.14 	& 7.16 	& 7.16 	& 7.09	& 6.59 	& 6.60 	& 6.70 	& 6.46 	& 6.33 \\
$ ^3A_{2u}$ 	& $1e_{1g}(\pi) \rightarrow 4e_{1u}(\sigma)$& \textit{6.52}	&		&		& 7.43	& 7.43	& 7.56	&		&		&		& 7.43	& 6.87	 &		&		& 6.92 &	\\
$ ^1A_{2u}$ 	& $1e_{1g}(\pi) \rightarrow 4e_{1u}(\sigma)$& \textit{6.54} & 6.55 	& 6.54 	& 7.53 	& 7.53 	& 7.63 	& 7.63 	& 7.65 	& 7.65 	& 7.53	& 7.01 	& 7.02 	& 7.17 	& 7.00 	& 6.93 \\ 
$ ^3E_{2u}$ 	& $1e_{1g}(\pi) \rightarrow 4e_{1u}(\sigma)$& \textit{6.54}	&		&		& 7.68	& 7.68	& 7.83	&		&		&		& 7.68	& 7.17	 &		&		& 7.06 &	\\
$ ^1E_{2u}$ 	& $1e_{1g}(\pi) \rightarrow 4e_{1u}(\sigma)$& \textit{6.55} & 6.55 	& 6.55 	& 7.71 	& 7.71 	& 7.84 	& 7.83 	& 7.84 	& 7.85 	& 7.71	& 7.21 	& 7.21 	& 7.32 	& 7.08 	& 6.95 \\
$ ^1A_{1u}$ 	& $1e_{1g}(\pi) \rightarrow 4e_{1u}(\sigma)$& \textit{6.59}	&		&		& 7.90	& 7.90	& 8.05	&		&		&		& 7.90	& 7.43	 &		&		& 7.18 &	\\
$ ^3A_{1u}$ 	& $1e_{1g}(\pi) \rightarrow 4e_{1u}(\sigma)$& \textit{6.59}	&		&		& 7.90	& 7.90	& 8.05	&		&		&		& 7.90	& 7.43	 &		&		& 7.19 &	\\
\hline                                                                                                                              
\multicolumn{17}{c}{Ionization threshold: $-\epsilon_{\text{HOMO}}$ (eV)} \\                                                                           
\hline                                                                                                                              
\multicolumn{2}{|c|}{                              }			& 6.50	&		&		& 9.72	&9.72		& 10.18	&		&		&		&9.72			& 9.15 &		&		&		&		\\
\hline
\multicolumn{17}{c}{ MAD of excitation energies with respect to EOM-CCSD (eV)} \\ 
\hline
Valence			&							& 0.39	&		&		& 0.32	& 0.22	& 0.29	&		&		&		& 0.43	& 0.93	 &		&		& - &\\ 
Rydberg 		&							& 0.49	&		& 		& 0.61	& 0.62	& 0.74	&		&		&		& 0.62	& 0.12	 &		&		& - &\\ 
Total 			&							& 0.45	&		& 		& 0.49	& 0.45	& 0.55	&		&		&		& 0.54	& 0.47	 &		&		& - &\\ 
\hline
\multicolumn{17}{c}{Oscillator strengths ($\times 10^{-2}$)} \\
\hline 
$ ^1E_{1u}$ 	& $1e_{1g}(\pi) \rightarrow 1e_{2u}(\pi^*)$ 	& 55.78 & 55.32 & 66.54 & 62.74 & 94.95 & 63.00 & 96.16 & 62.57 & 95.89 & 62.42	& 71.49 & 53.89 & 49.71 & 66.41 & 125,120,88, \\
$ ^1A_{2u}$ 	& $1e_{1g}(\pi) \rightarrow 4e_{1u}(\sigma)$	& 2.11 	& 1.71 	& 2.40 	& 7.10 	& 7.55 	& 8.27 	& 8.82 	& 7.81 	& 8.28 	& 6.87	& 7.69 	& 8.22 	& 11.36 & 7.04 	&- \\
\hline 
\end{tabular}
}
\caption{Excitation energies and oscillator strengths of C$_6$H$_6$ calculated by linear-response HF and KS (with the LDA functional), by the linear-response range-separated methods RSH, RSH-TDA, LC, and RSH-{\fHx} (with the short-range LDA functional and $\mu=0.4$ bohr$^{-1}$), and by EOM-CCSD taken as reference, using the Sadlej+ basis set. Excitation energies above the ionization threshold are indicated in italics. Except for the E$_{1g}$ states whose assignments were trivial, all the other states correspond to the orbital transitions $e_{1g} \fl e_{2u}$ and $e_{1g} \fl e_{1u}$, which lead to $B_{1u}\oplus E_{1u} \oplus B_{2u}$ and $A_{1u}\oplus E_{2u} \oplus A_{2u}$ manifolds, respectively, in the $\mathcal{D}_{6h}$ symmetry point-group. Since the calculations were performed in the $\mathcal{D}_{2h}$ subgroup, some symmetry information for these states was lost but the assignment could be done using the degeneracy of the states.}
\label{C6H6table}
\end{table*}
\endgroup

\end{document}